          [2001/04/25 1.1 (PWD)]
\documentclass[a4paper,twocolumn]{esapub2005} 
\usepackage{times}
\usepackage{natbib}
\usepackage{graphicx}

\usepackage{color}

\def\farcs{\hbox{$.\!\!^{\prime\prime}$}}
\def\arcsec{\hbox{$^{\prime\prime}$\,}}

\title{Simulation and analysis of VIM measurements: feedback on design parameters}
\author[1]{D.\ Orozco Su\'arez}
\author[1]{L.R.\ Bellot Rubio}
\author[2]{S.\ Vargas}
\author[2]{J.A.\ Bonet}
\author[2]{V.\ Mart\'{\i}nez Pillet}
\author[1]{J.C.\ del Toro Iniesta}
\affil[1]{Instituto de Astrof\'isica de Andaluc\'ia (CSIC), Apdo.\ 3004, 18080 Granada, Spain}
\affil[2]{Instituto de Astrof\'isica de Canarias, 38205 La Laguna, Tenerife, Spain}

\begin{document}


\maketitle

\begin{abstract}
The Visible-light Imager and Magnetograph (VIM) proposed for the ESA
Solar Orbiter mission will observe a photospheric spectral line at
high spatial resolution. Here we simulate and interpret VIM
measurements.  Realistic MHD models are used to synthesize "observed"
Stokes profiles of the photospheric Fe {\small I} 617.3 nm line. The
profiles are degraded by telescope diffraction and detector pixel size
to a spatial resolution of 162~km on the solar surface.  We study the
influence of spectral resolving power, noise, and limited wavelength
sampling on the vector magnetic fields and line-of-sight velocities
derived from Milne-Eddington inversions of the simulated measurements.
VIM will provide reasonably accurate values of the atmospheric
parameters even with filter widths of 120~m{\AA} and 3 wavelength
positions plus continuum, as long as the noise level is kept below
$10^{-3} I_{\rm c}$.

\end{abstract}

\section{Introduction}

The goal of the Visible-light Imager and Magnetograph (VIM; Marsch et
al.~2005), one of the remote sensing instruments onboard Solar
Orbiter, is to obtain high resolution maps of the vector magnetic
field and the line-of-sight (LOS) velocity in the solar photosphere.
This information is essential to understand not only the physical
processes occurring there, but also the magnetic coupling of the
different atmospheric layers. In addition, VIM will carry out 
local and global helioseismic studies of the Sun.

The inference of LOS velocity and vector magnetic field (strength,
inclination and azimuth) maps, commonly called Dopplergrams and vector
magnetograms, requires the observation and subsequent analysis of a
spectral line in polarized light. The atmospheric parameters (physical
quantities) are retrieved from the polarimetric measurements by
techniques based on either the radiative transfer equation or look-up
tables (Bellot Rubio 2006). VIM consists of two telescopes: the High
Resolution Telescope (VIM-HRT) and the Full Disk Telescope
(VIM-FDT). Spectropolarimetry is carried out using a double
Fabry-P\'erot interferometer, conceptually based on LiNbO$_3$ etalons,
which performs the wavelength selection within the spectral line, and
two polarization modulation packages, based on liquid crystal
retarders, to modulate the polarization of the incident light.
Solanki et al.~(2006) describe the main properties of the instrument
and a possible configuration for VIM on Solar Orbiter. The
photospheric line to be observed is Fe~{\small I}~617.3~nm.

The purpose of the present work is to investigate how well we are able
to infer atmospheric parameters from VIM-HRT data, providing feedback
to optimize its design. In many respects, VIM-HRT is very similar to
the Imaging MAgnetograph eXperiment (IMaX; Mart\'{\i}nez Pillet et
al.~2004), the vector polarimeter of the SUNRISE balloon mission
(Gandorfer et al.~2006). We have carried out extensive tests to
improve the SUNRISE/IMaX performance (e.g.\ Orozco Su\'arez et al.\
2006).  Here we use this experience to study the influence of spectral
resolution and wavelength sampling on the accuracy of the atmospheric
parameters derived from VIM-HRT measurements. We show that filter
widths of 120~m\AA\/ and 4 wavelength samples (3 across the line and
one in the continuum) would allow VIM to achieve its science goals.

\section{Methodology}

We simulate the observational process of VIM, from the measurement of
spectra to the determination of physical quantities, as follows. 
First, we use model atmospheres
that describe the Sun in the more realistic way possible. These
atmospheres allow us to simulate {\em observations} by synthesizing
the Stokes profiles ($I$, $Q$, $U$, $V$) of Fe~{\small
I}~617.3~nm. The polarization signals are spatially degraded
considering telescope diffraction and detector pixel size. We also
degrade the profiles applying a spectral PSF, add noise, and select a
few wavelength samples across the line. The simulated ``observations''
are then analyzed by means of inversion techniques. Comparing the
retrieved parameters with the real ones we estimate the uncertainties
of the inferences.

VIM-HRT is expected to observe the solar photosphere at resolutions of
$\sim$\,150~km. No ground-based telescope has ever provided
spectropolarimetric measurements at such a resolution. For this
reason, the atmospheres needed to synthesize the Stokes profiles are
taken from MHD simulations (V\"ogler et al.~2005; Sch\"ussler et
al.~2003). More specifically, we use a simulation run representing a
quiet Sun area with mixed-polarity magnetic fields and unsigned
average flux of $\sim$\,150~G. The duration of the simulation sequence
is roughly 5 minutes with a cadence of 10 seconds.  The horizontal and
vertical extents of the computational box are 6 and 1.4~Mm,
respectively. The synthesis of Stokes profiles is carried out using
the SIR code (Ruiz Cobo \& del Toro Iniesta 1992).  The line
considered here, Fe~{\small I} 617.3~nm, is sampled at 61 wavelength
positions in steps of 1~pm. The atomic parameters have been taken
from the VALD database (Piskunov et al.~1995).  These Stokes profiles
represent the {\em real} Sun. To determine the atmospheric parameters
from them we use a least-square inversion technique based on
Milne-Eddington (ME) atmospheres.

\begin{figure}[!t]
\centering
\resizebox{\hsize}{!}{\includegraphics[trim=4mm 8mm 0mm 12mm,clip]{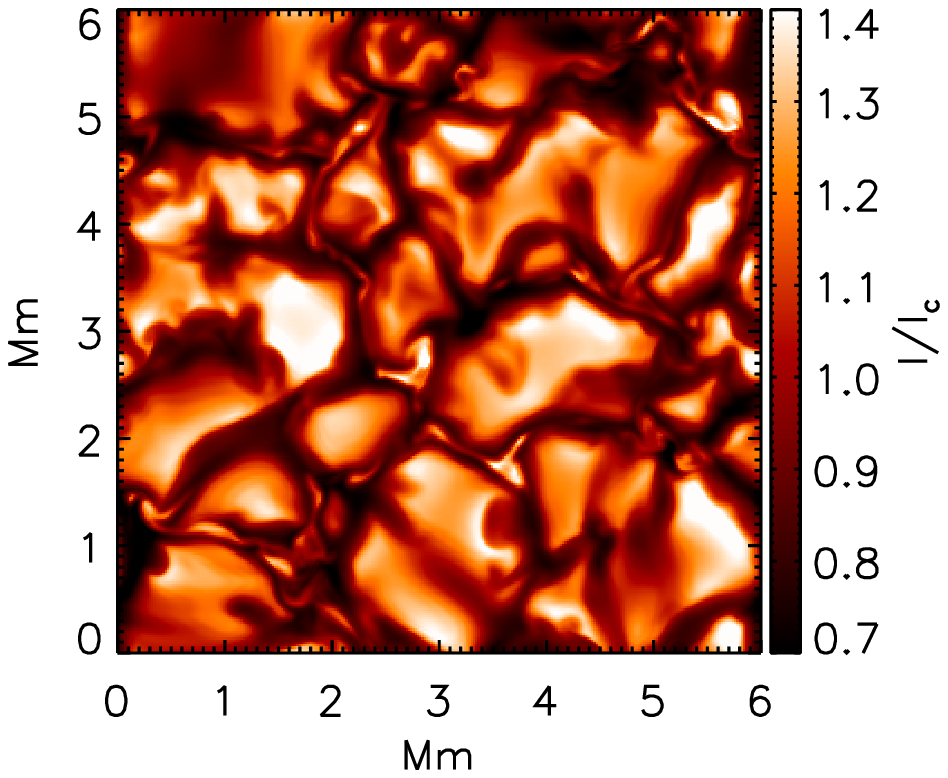}}
\resizebox{\hsize}{!}{\includegraphics[trim=4mm 8mm 0mm 12mm,clip]{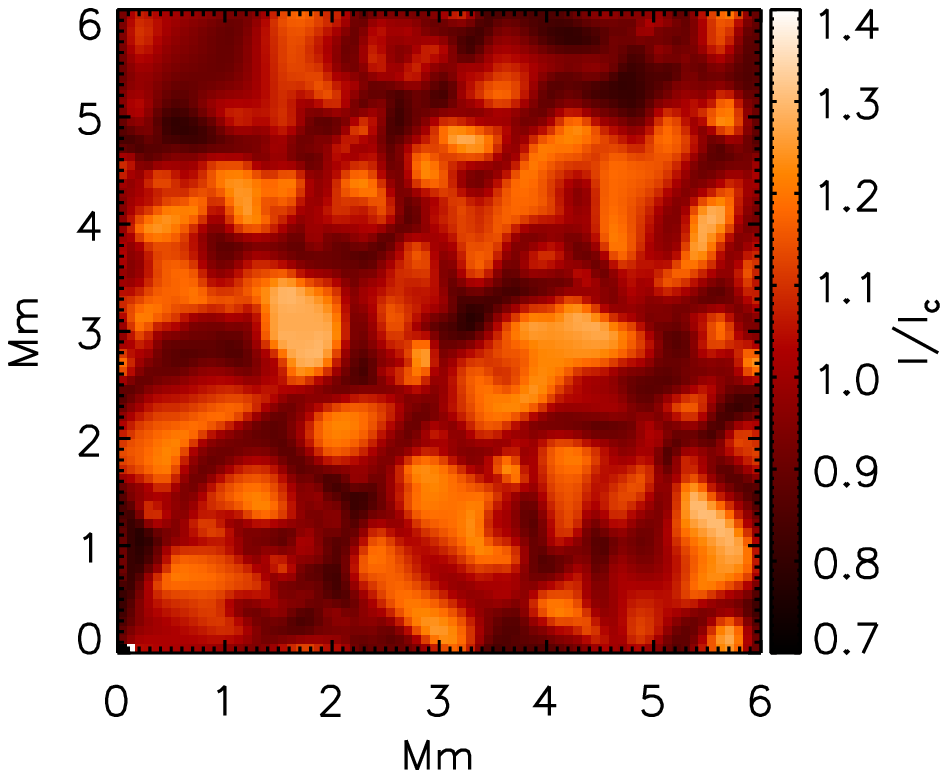}}
\caption{Maps of the normalized continuum intensity for the non-degraded data 
(top) and for the spatially degraded data (bottom) considering telescope 
diffraction and pixel size. Note that the color scales are the same in 
the two maps. 
\label{fig:continuo}}
\end{figure}

\begin{figure}[!t]
\centering
\resizebox{.95\hsize}{!}{\includegraphics[bb=74 360 578 720]{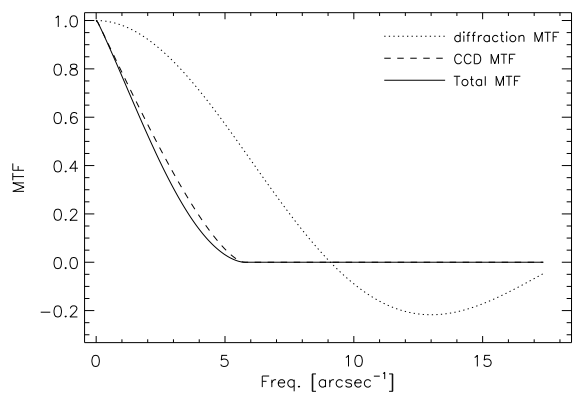}}
\caption{Dotted line: MTF of the CCD (pixelation effect); dashed line:
diffraction limited MTF; and solid line: MTF combining both
effects. \label{MTF}}
\end{figure}

The sampling interval in the MHD simulations is 0\farcs0287 (grid
resolution), implying a spatial resolution of 0\farcs057 or 41.6~km on
the solar surface. The spatial resolution provided by the aperture of
VIM\footnote{VIM-HRT is equivalent to a 0.73m telescope at 1 AU}
operating at 617.3 nm is $\sim$0\farcs17 (i.e., $\sim$127 km on the
Sun), but the sampling interval (0\farcs11) imposed by the CCD limits
the spatial resolution to 162~km ($\sim$0\farcs22) on the Sun.
Thus, in order to properly simulate VIM observations, the data images 
derived from the model (the synthetic Stokes profiles) have to be 
spatially degraded by telescope diffraction and detector pixel size. 
Figure \ref{fig:continuo} shows maps of the normalized continuum 
intensity for the original data (the theoretical model) and for 
the spatially degraded data. The main visible effect of the 
degradation process is the loss of contrast from $\sim$14\% to 
$\sim$11\% in the continuum. The CCD grid and the disappearance of 
small scale structures are also very noticeable. Figure~\ref{MTF} 
shows the MTFs representing, in the Fourier domain, the filtering 
of spectral components induced by telescope diffraction and 
pixelation effects in the CCD.

The most favorable (ideal) case is one in which the instrument
measures the spatially degraded Stokes profiles with no noise, very
high spectral resolution, and critical wavelength sampling. Inversion
techniques would be able to infer correct atmospheric parameters from
this kind of observations, but one needs complex model atmospheres
with vertical gradients to describe the height variation of the
physical quantities within the same pixel. Such models are not
feasible because of the limited data processing capabilities onboard
Solar Orbiter. Thus, ME inversions represent the best option to
interpret VIM-HRT measurements: they do not retrieve stratifications,
but are simple and often provide reasonable averages of the physical
quantities over the line formation region (Westendorp Plaza et
al.~2001; Bellot Rubio 2006).

In the present work we consider the results of ME inversions of 
the spatially degraded Stokes profiles with no noise, no spectral 
PSF, and 61 wavelength samples as {\em the reference} solution. By 
comparing this reference with the outcome of ME inversions of 
the same Stokes profiles affected by noise, limited spectral 
resolution, and wavelength sampling, we quantify the loss of 
information induced by the measuring process, avoiding errors 
due to the ME assumption.


\begin{figure}[!t]
\centering
\resizebox{.88\hsize}{!}{\includegraphics{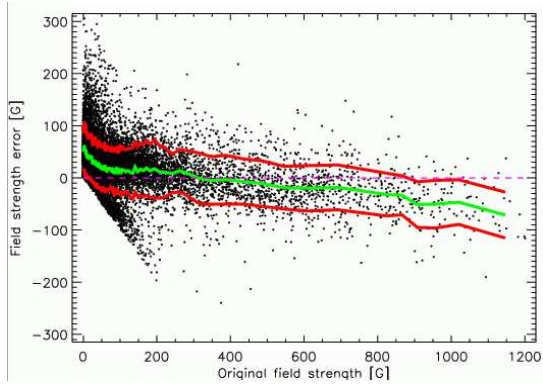}}
\caption{Field strength errors for a 120 m{\AA} instrumental
width and noise at the level of $10^{-3}$. The green and red 
lines represent the mean and rms errors, respectively.
\label{individuales}}
\end{figure}

\section{Test results}

VIM uses a Fabry-P\'erot interferometer to perform the wavelength
selection within the line. The finite spectral resolution of the
instrument reduces the amount of information carried by the line, and
therefore is a source of uncertainties in the determination of
atmospheric parameters. The spectral PSF of VIM can be described as 
a Gaussian function whose FWHM lies somewhere between 75 and 120~m{\AA}.

We estimate the effect of limited spectral resolution as follows.  The
synthetic Stokes profiles are convolved with PSFs of different
widths. Specifically, we vary the FWHM from zero to 200~m{\AA} in
steps of 10~m{\AA}. We then add noise at the level of $10^{-3}$, apply
a ME inversion to the profiles sampled at 61 wavelength positions, and
compare the inferred maps with our {\em reference}. The ME inversion
process determines 9 free parameters. The magnetic filling factor is
fixed to unity and no stray light is considered. We use the same
initial guess model for all inversions, allowing a maximum of 300
iterations. Three simulation snapshots (17700 pixels) have
been inverted in this way.

To analyze the test results we calculate the mean and rms values of
the errors (defined as the difference between the inferred and the
{\em reference} parameters). As an example, Fig.~\ref{individuales} 
shows the field strength errors resulting from the inversion of the 
Stokes profiles convolved with a 120~m{\AA} FWHM filter. Each point 
represents an individual pixel. The solid lines give the mean and 
rms errors. 

\begin{figure}[!t]
\centering
\resizebox{0.88\hsize}{!}{\includegraphics[trim=5mm 0mm 2mm 0mm,clip]{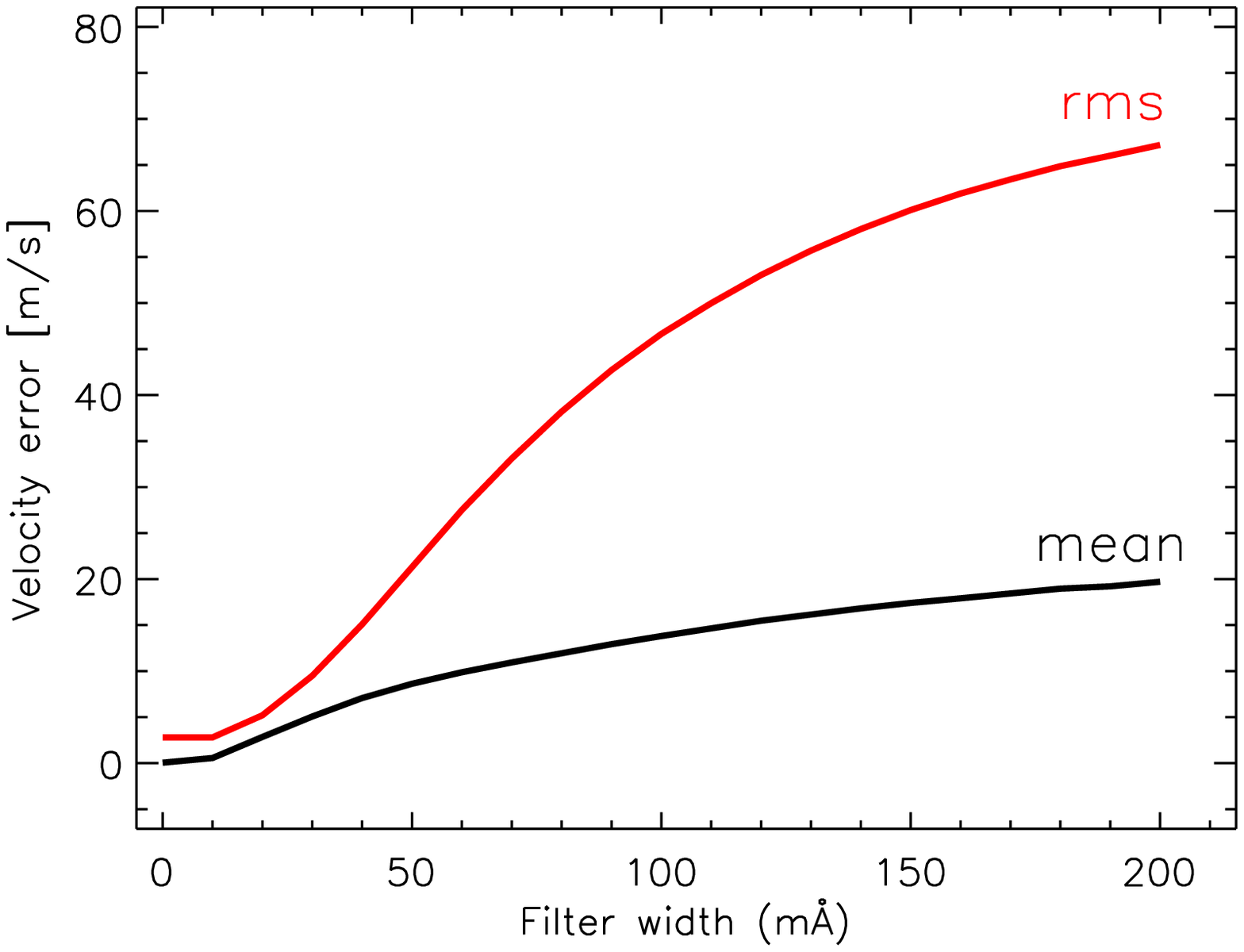}}
\resizebox{0.88\hsize}{!}{\includegraphics[trim=5mm 0mm 2mm 0mm,clip]{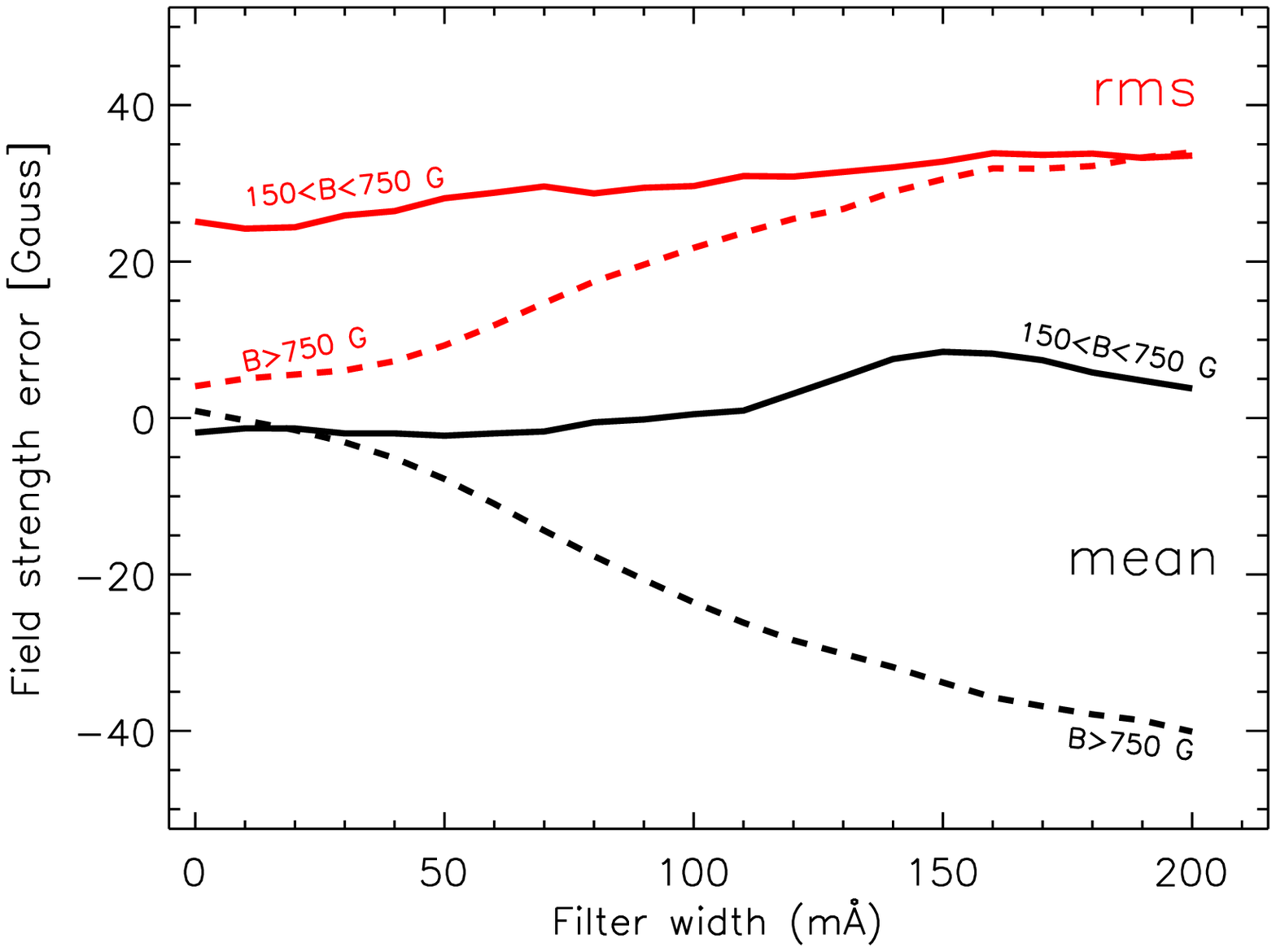}}
\resizebox{0.88\hsize}{!}{\includegraphics[trim=5mm 0mm 2mm 0mm,clip]{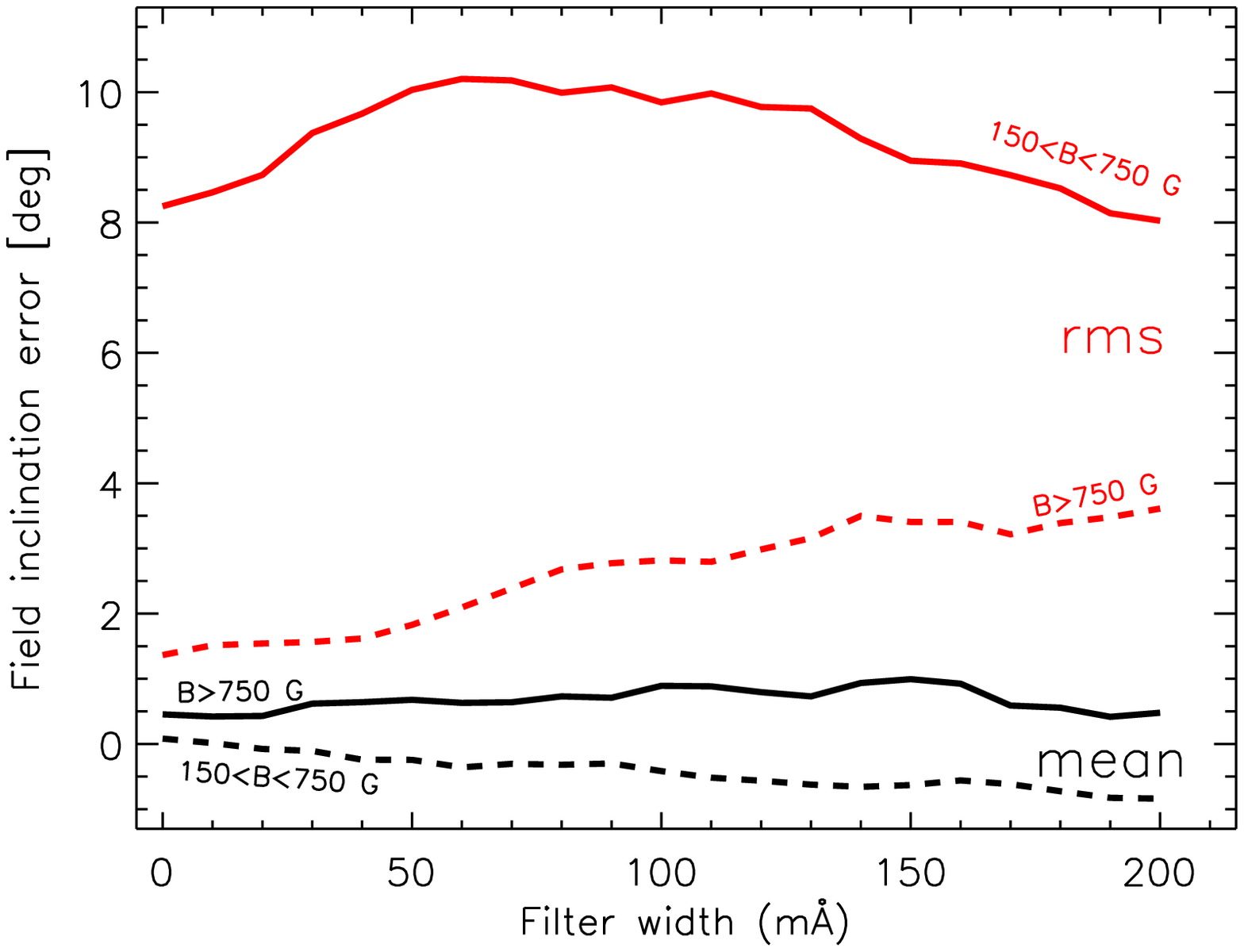}}
\caption{Variation of the mean (black) and rms (red) values of 
the error with the FWHM of the instrumental PSF. Top: LOS velocity.
Middle: magnetic field strength. Bottom: magnetic field inclination.
The dashed curve in the central and bottom panels represents pixels
with field strengths larger than 750~G (in the reference map) and the
solid curve field strengths ranging from 150 to 750~G.\label{fig:filtro1}}
\end{figure}

Figure \ref{fig:filtro1} shows the variation of the mean and rms
errors with the FWHM, for the LOS velocity (upper panel) and the
magnetic field strength and inclination (middle and bottom panels).
In the last two panels we have considered only pixels whose Stokes
$Q$, $U$ {\em or} $V$ amplitudes exceed three times the noise
level. Different conclusions can be drawn from this figure. First, 
we note that the rms errors for filter widths of 0~m\AA\/ are
$\sigma_{\rm v} \sim 4$~m/s in velocity, $\sigma_{\rm B} \leq 30$~G 
in field strength, and $\sigma_\gamma \leq 8^{\rm o}$ in field 
inclination. These errors are solely due to the photon noise of 
$10^{-3}$ added to the observables (which was zero in the reference 
profiles). Therefore, {\em they represent the minimum uncertainties 
that VIM would produce} even if the spectral line is critically 
sampled at 61 wavelength positions. 

The mean and rms errors of the velocity increase with filter width,
although the variation is weak. We estimate rms errors of about 30~m/s
and 50~m/s for 60~m{\AA} and 120~m{\AA} filter widths, respectively.
The errors in the magnetic field strength also vary smoothly with the
FWHM.  For filters narrower than 120~m{\AA}, the rms errors are always
smaller than $\sim$30~G. Interestingly, the mean errors increase with
increasing field strength: in the range 150--750~G they are roughly
constant, which is not the case for fields stronger than 750~G.  The
fact that the mean errors of field strength and velocity vary with the
FWHM is related to the asymmetries of the profiles.  Stokes profiles
formed in real atmospheres exhibit asymmetries induced by vertical
gradients of the atmospheric parameters. The Stokes profiles coming
from ME atmospheres are {\em symmetric}, however. While the spectral
PSF smears out the asymmetries, it also allows better fits to the
observations. Consequently, the mean errors vary with filter width,
and the variation is larger for stronger fields.

\begin{figure}[!t]
\centering
\resizebox{0.82\hsize}{!}{\includegraphics[trim=5mm 0mm 2mm 0mm,clip]{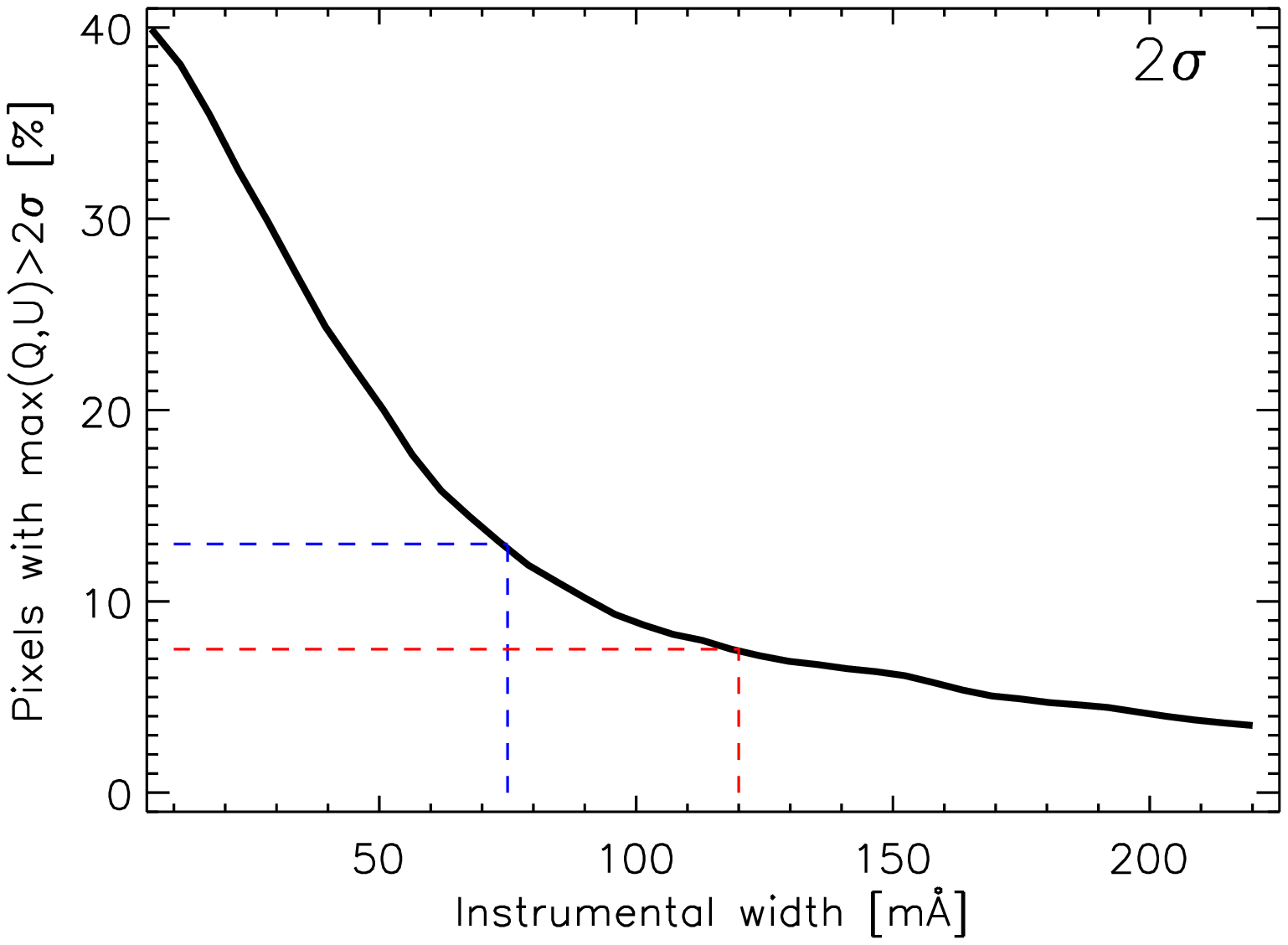}}
\resizebox{0.82\hsize}{!}{\includegraphics[trim=5mm 0mm 2mm 0mm,clip]{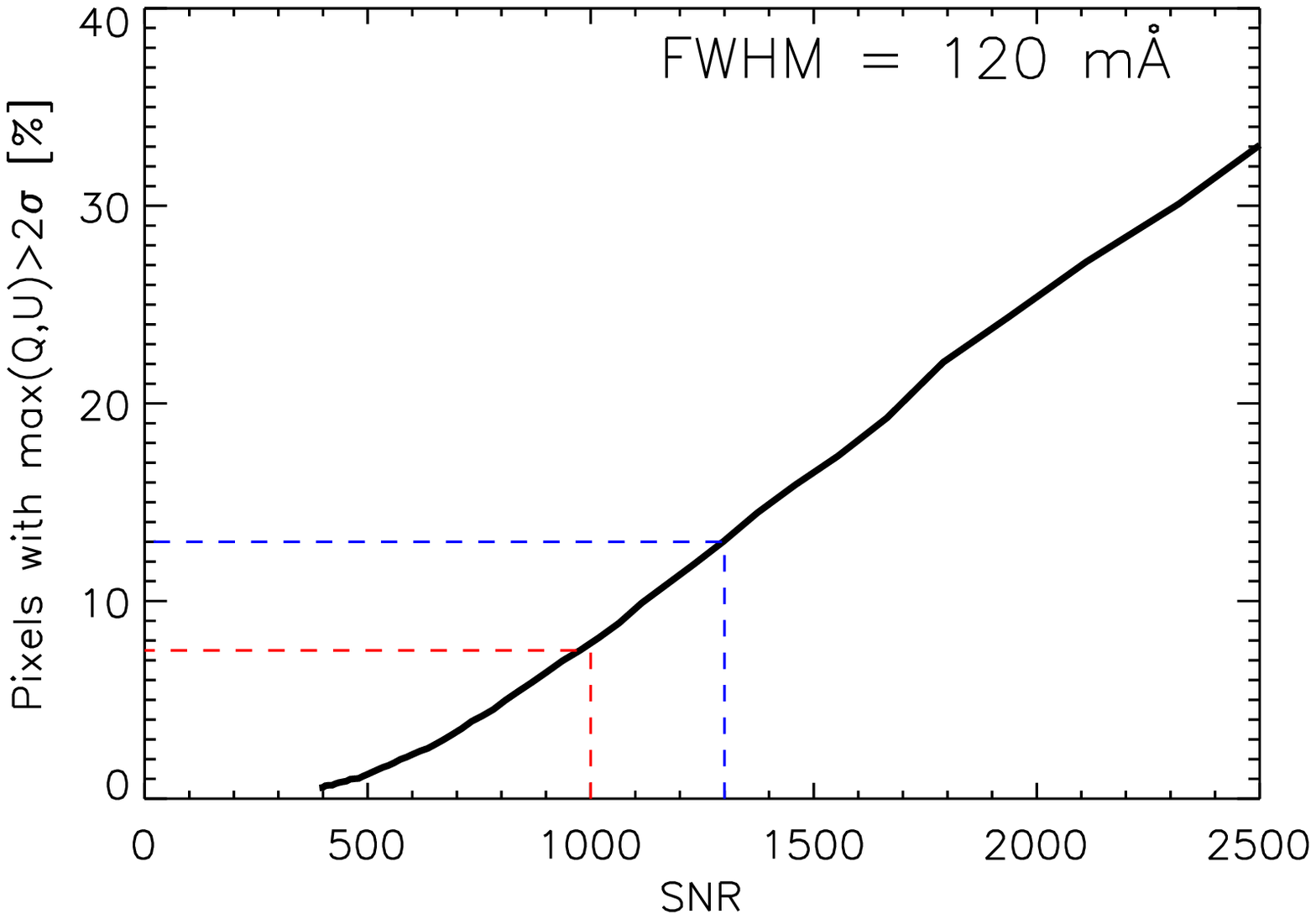}}
\caption{Variation of the number of pixels whose Stokes $Q$ or $U$ 
amplitudes exceed twofold the noise level, in percent, as a function
of the instrumental filter width (top) and as a function of the
signal-to-noise ratio for a filter width of 120~m\AA\/ (bottom).
\label{fig:filtro2}}
\end{figure}

From this analysis we conclude that instrumental profiles of up to
120~m{\AA}~FWHM provide accurate results. It is important to keep in
mind, however, that the spectral PSF also affects the Stokes profiles
in two different ways: first it reduces their amplitudes, and second
it smooths the asymmetries out. To quantify the first effect, the
upper panel of Fig.~\ref{fig:filtro2} shows, as a function of filter
width, the percentage of pixels whose Stokes $Q$ or $U$ amplitudes
exceed twofold the noise level. This percentage decreases rapidly with
the FWHM of the PSF. In other words: {\em the ability to detect linear
polarization signals strongly depends on the instrumental profile}, at
least in quiet Sun regions. For instance, 6\% of the pixels are no
longer detectable in linear polarization when the filter width is
increased from 75 to 120~m{\AA}. This loss of sensitivity can be
compensated by lowering the noise level. In the bottom panel of
Fig.~\ref{fig:filtro2}, we represent the percentage of pixels with
detectable linear polarization signals against the signal-to-noise
ratio (SNR) for a fixed filter width of 120~m{\AA}. The variation is
almost linear. If we are to recover the previous loss of 6\% of the
pixels, the SNR has to be increased from 1000 to 1300. This translates
into a 1.7 factor in exposure time, which may have some unwanted
consequences on high spatial resolution observations.

VIM-HRT will achieve a spatial resolution of about 150~km in the solar
photosphere. At this resolution the smallest dynamical structures
accessible evolve on time scales of 10--50 seconds (assuming a scale
height of 100 km). Thus, the scanning of the spectral line should not
take longer. This fact limits VIM to sample only a few wavelength
positions within the line. Currently, scans of five wavelength
positions plus one in the nearby continuum are being considered.  The
limited wavelength sampling introduces additional uncertainties in the
inference process. To determine these errors we carry out ME
inversions of the Stokes profiles sampled with different numbers of
wavelength points, from 2 to 8, plus the continuum. First, the
profiles have been convolved with a 120~m{\AA}~FWHM filter and have
been added noise at the level of $10^{-3} I_{\rm c}$. The inversion 
is carried out in the same conditions as before. Again, we compare 
the inferred maps with the {\em reference} solution.

\begin{figure}[!t]
\centering
\resizebox{0.8\hsize}{!}{\includegraphics[trim=5mm 0mm 2mm 0mm,clip]{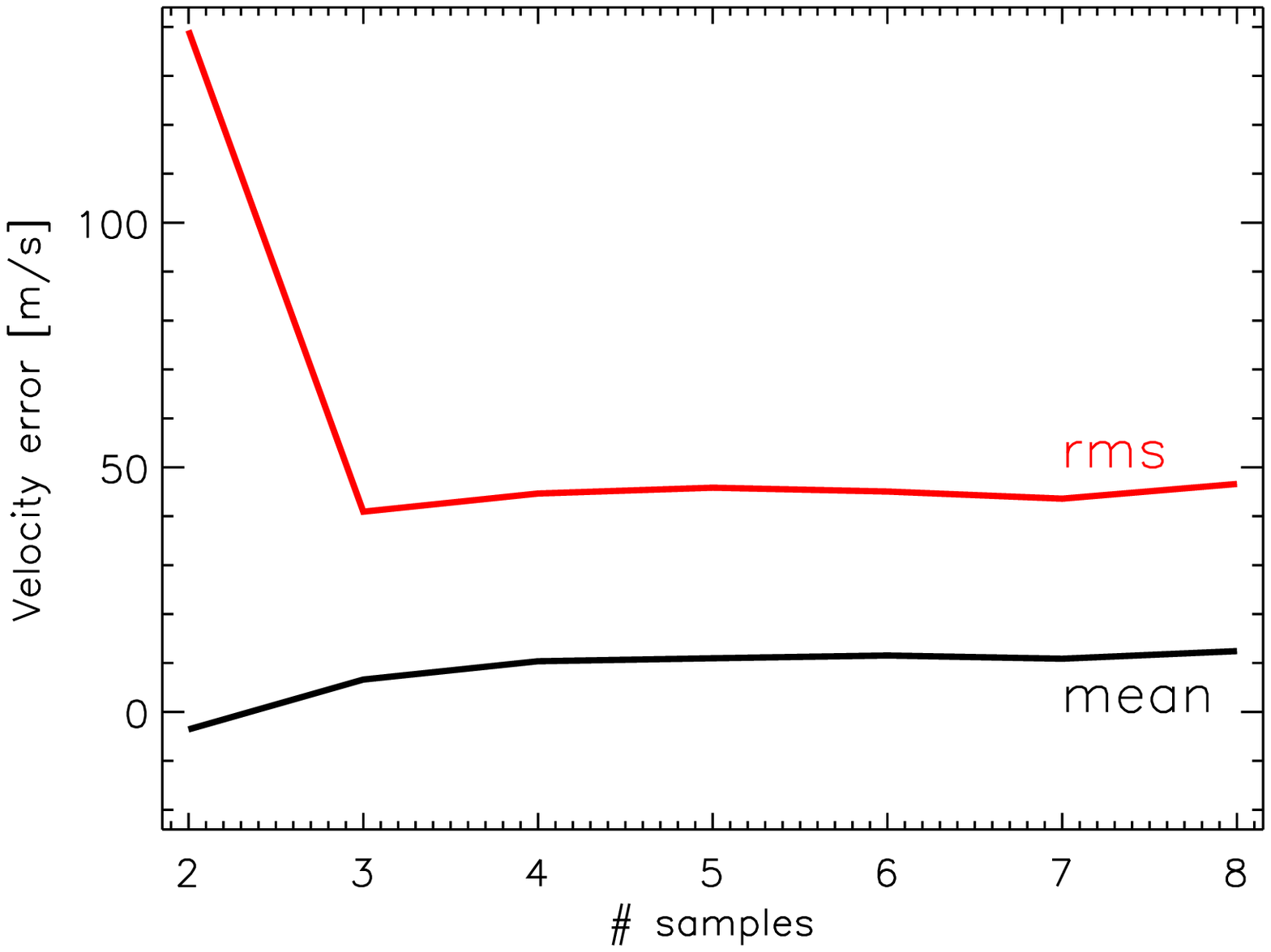}}
\resizebox{0.8\hsize}{!}{\includegraphics[trim=5mm 0mm 2mm 0mm,clip]{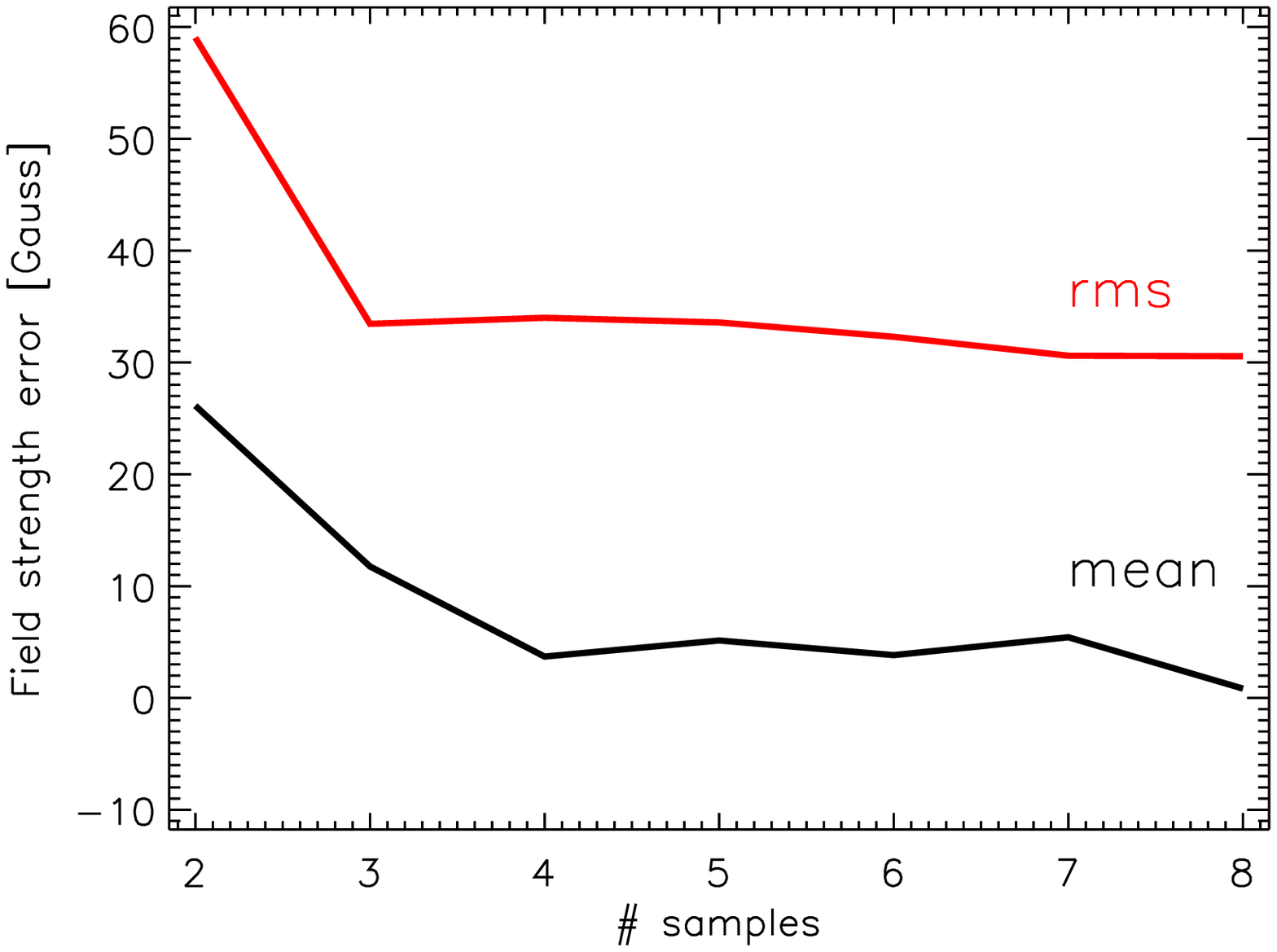}}
\resizebox{0.8\hsize}{!}{\includegraphics[trim=5mm 0mm 2mm 0mm,clip]{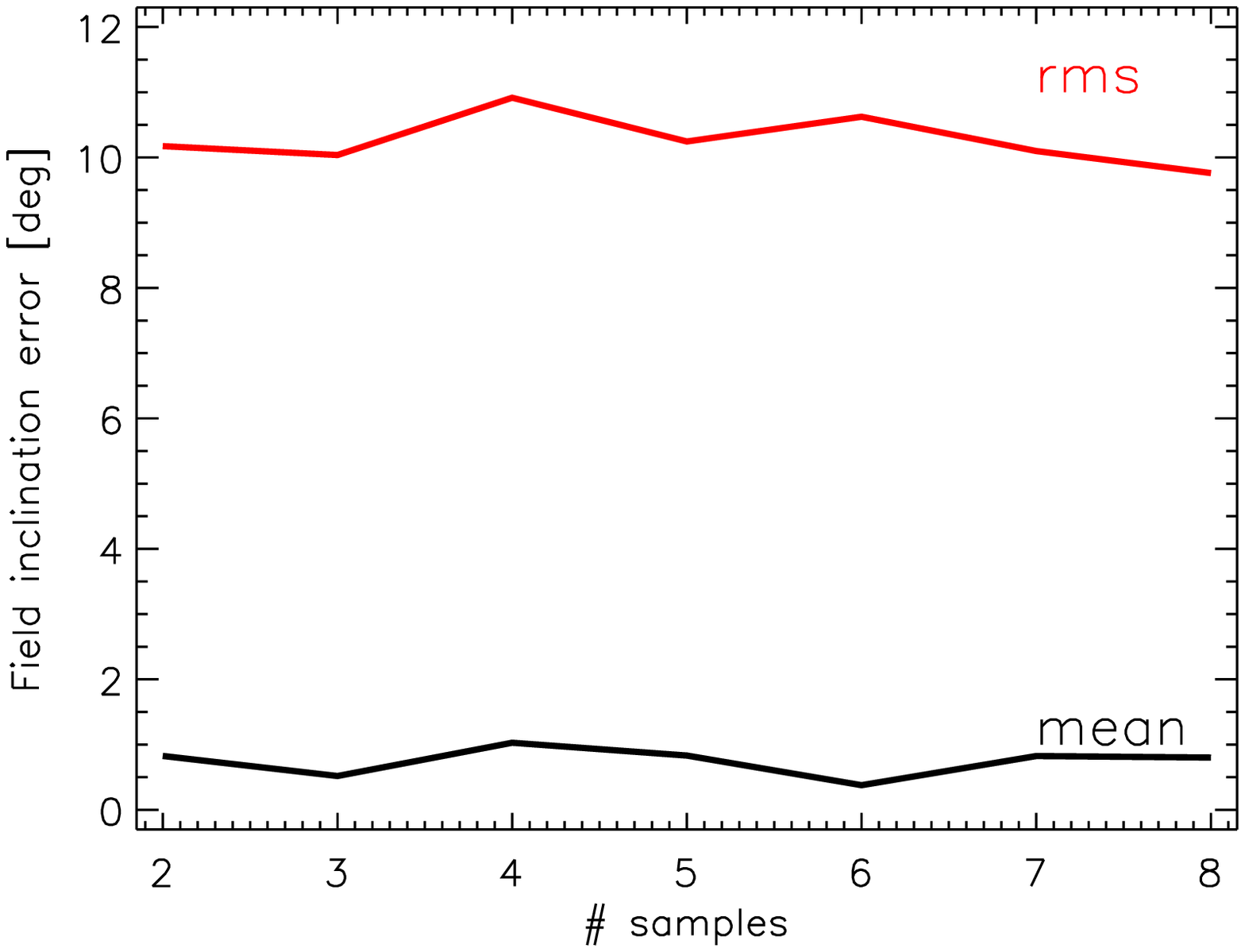}}
\caption{Variation of the mean (black) and rms (red) errors in 
LOS velocity (top), field strength (middle), and field inclination
(bottom) with the number of wavelength samples, for a filter width of
120~m\AA\/. The horizontal axis indicates the number of samples 
within the spectral line.
\label{fig:landas}}
\end{figure}

Figure \ref{fig:landas} shows the variation of the mean and rms errors
with the number of wavelength samples, for the LOS velocity (top), 
field strength (middle), and field inclination (bottom). Only pixels 
whose Stokes $Q$, $U$ or $V$ amplitudes are larger than three times 
the noise level have been considered for the magnetic parameters. The 
results are somewhat surprising. We find that the mean and rms errors
of the velocity do not change much with the number of samples if the
line is observed in at least three wavelength positions. In that case, 
the mean and rms errors are about 10 and 50~m/s, respectively. The 
field strength and field inclination errors do not change either with
the number of samples, provided it is larger than 3. The reason for
such a behavior is the strong smearing of the Stokes profiles after
the instrument action. No conspicuous details remain that can be
detected by five or six samples better than by just three. It is 
important to remark, however, that the sampling will further reduce 
the number of detectable profiles over those shown in 
Fig.~\ref{fig:filtro2}, since in general the observed wavelength
positions will not coincide with the maximum Stokes $Q$ or $U$ signals.

\begin{figure*}[!ht]
\centering
\resizebox{0.246\hsize}{!}{\includegraphics[trim=25mm 3mm 6mm
0mm,clip]{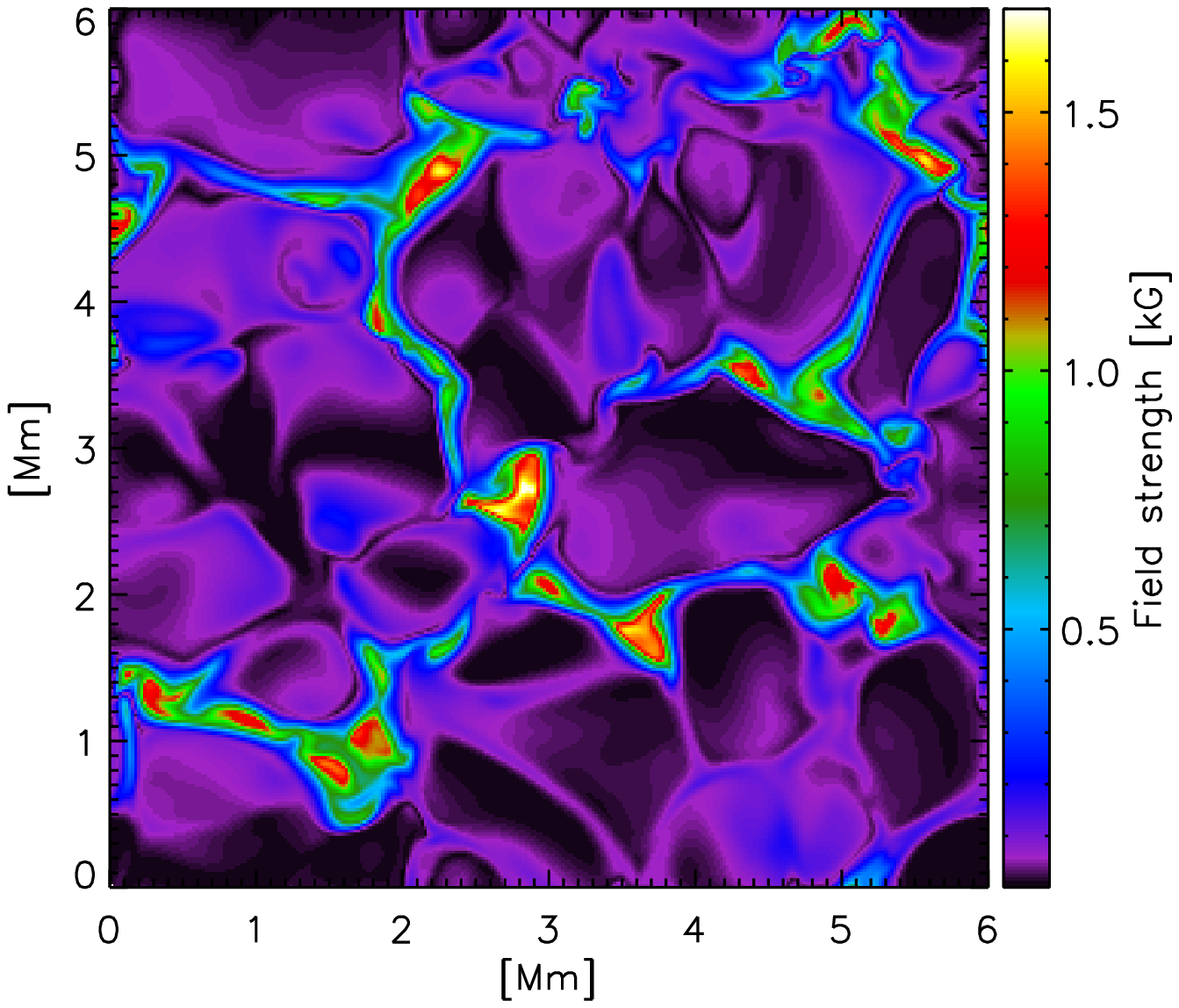}}
\resizebox{0.246\hsize}{!}{\includegraphics[trim=25mm 3mm 6mm
0mm,clip]{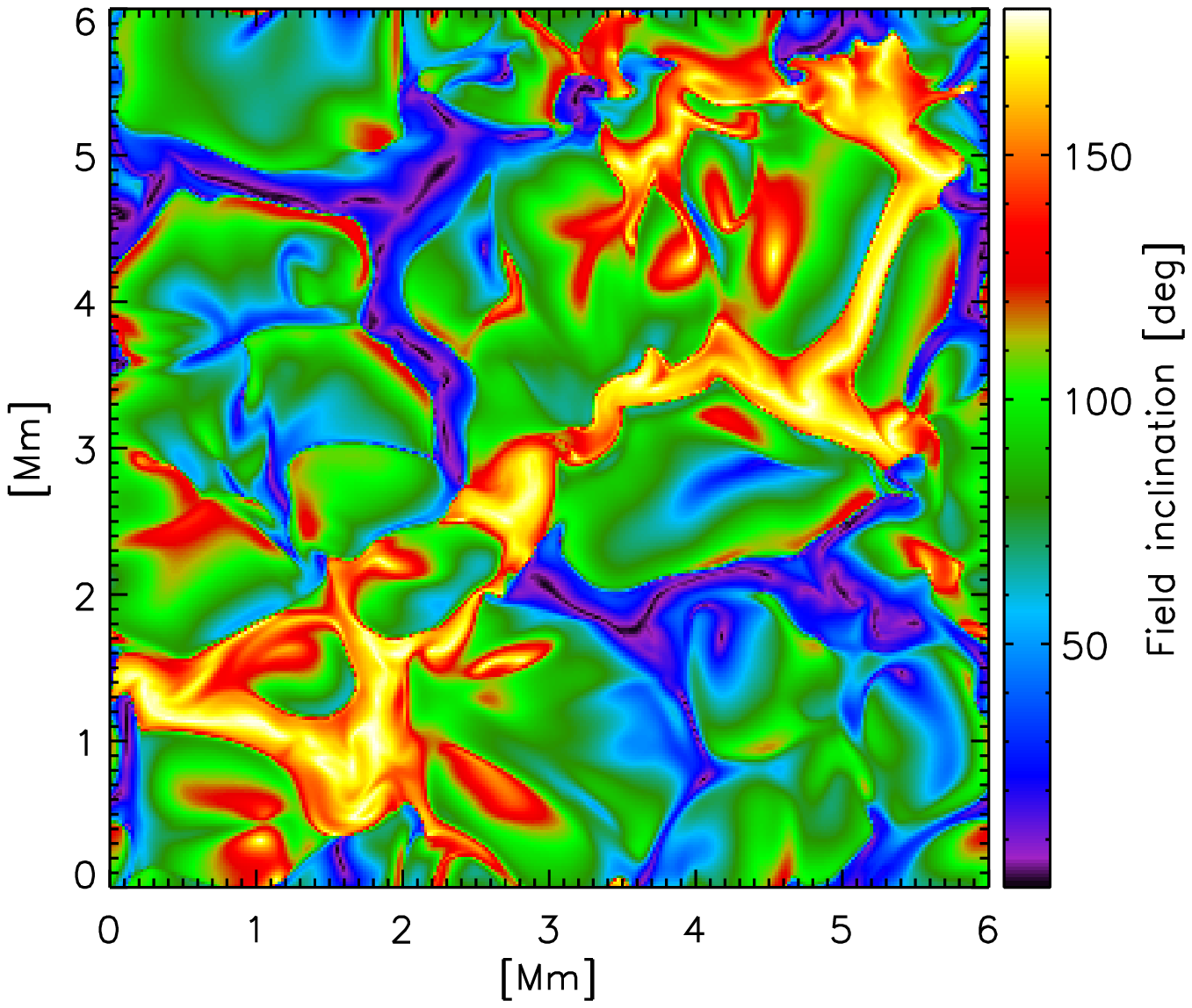}}
\resizebox{0.246\hsize}{!}{\includegraphics[trim=25mm 3mm 6mm
0mm,clip]{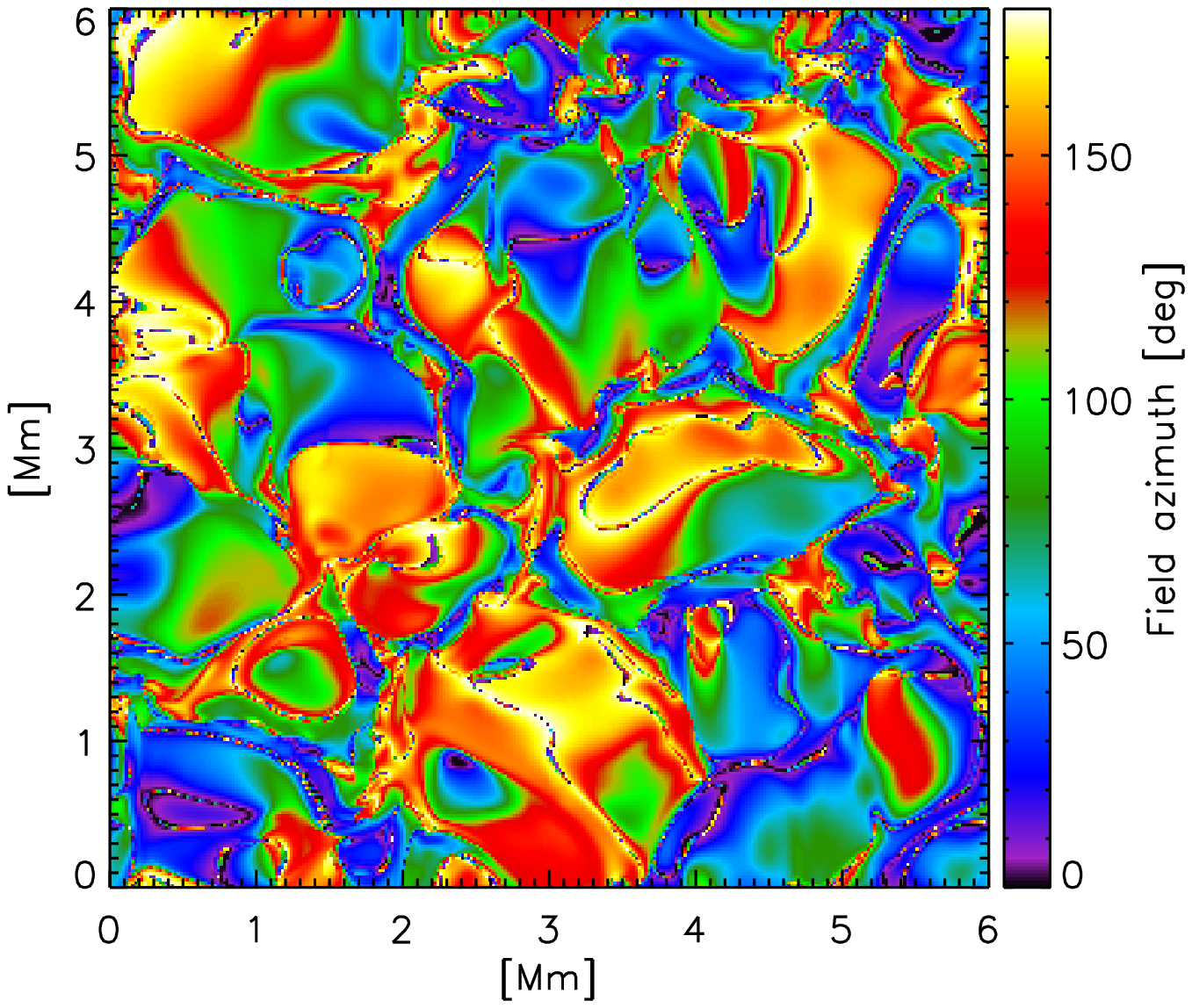}}
\resizebox{0.246\hsize}{!}{\includegraphics[trim=25mm 3mm 6mm
0mm,clip]{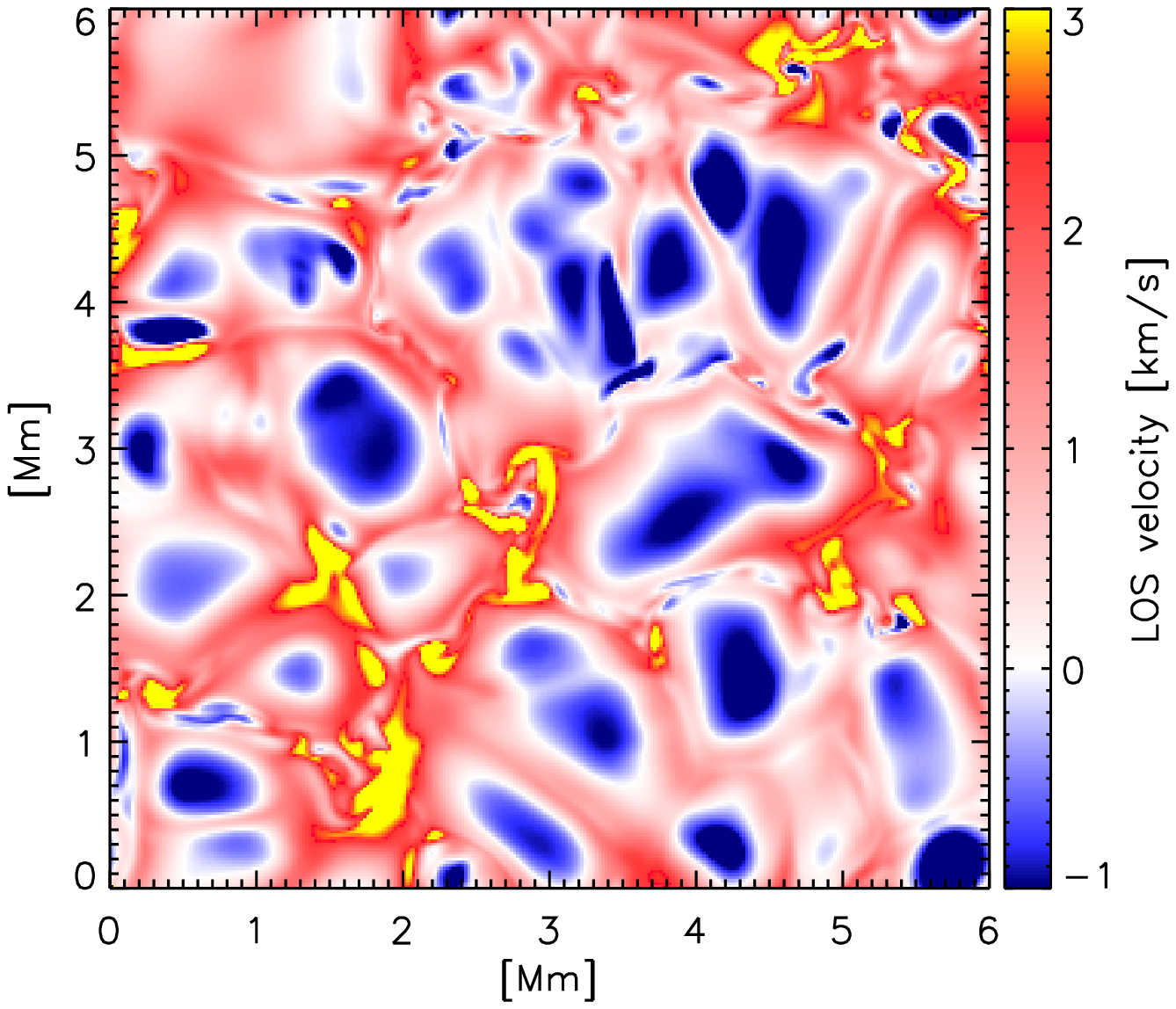}}
\resizebox{0.246\hsize}{!}{\includegraphics[trim=25mm 3mm 6mm
0mm,clip]{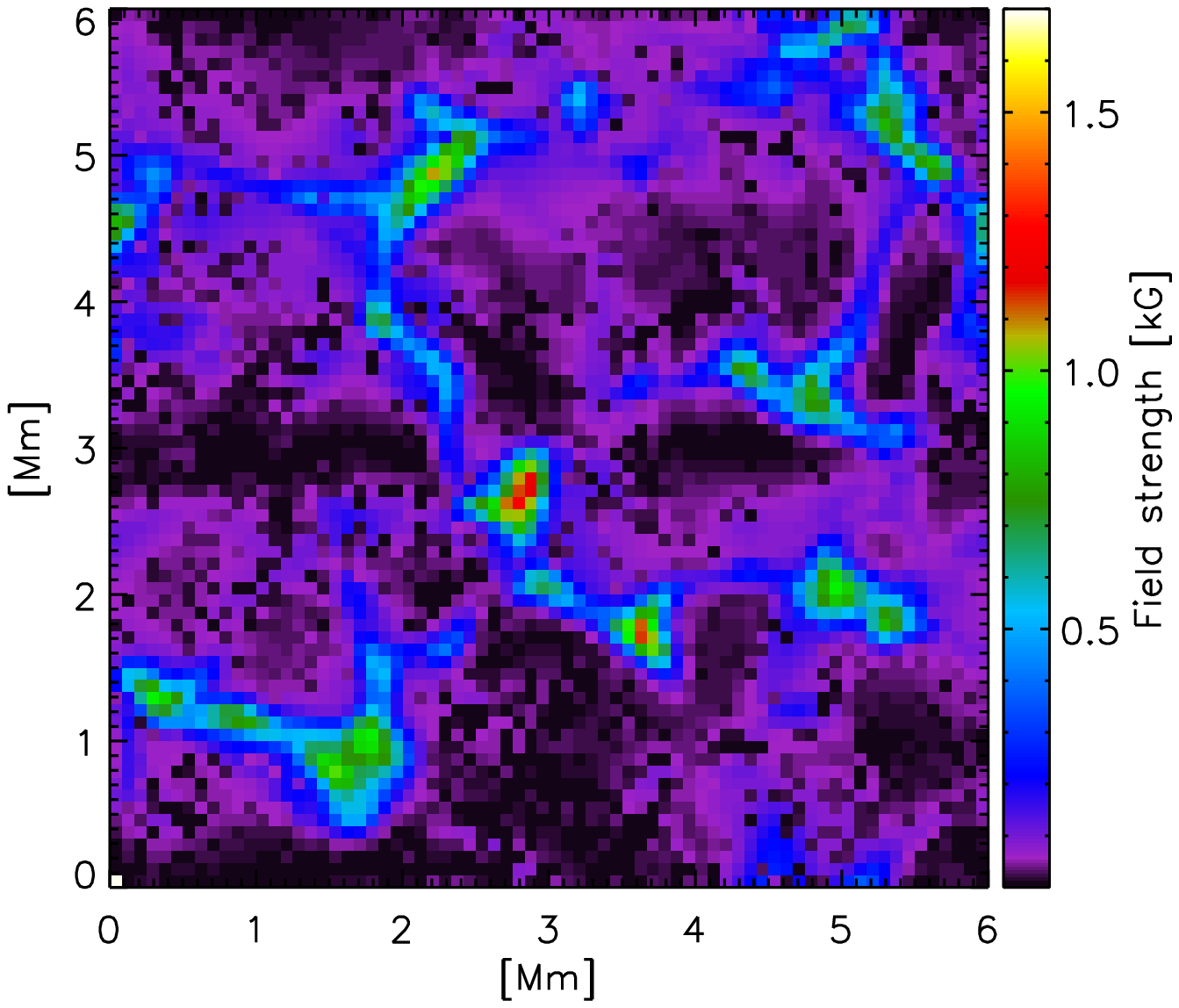}}
\resizebox{0.246\hsize}{!}{\includegraphics[trim=25mm 3mm 6mm
0mm,clip]{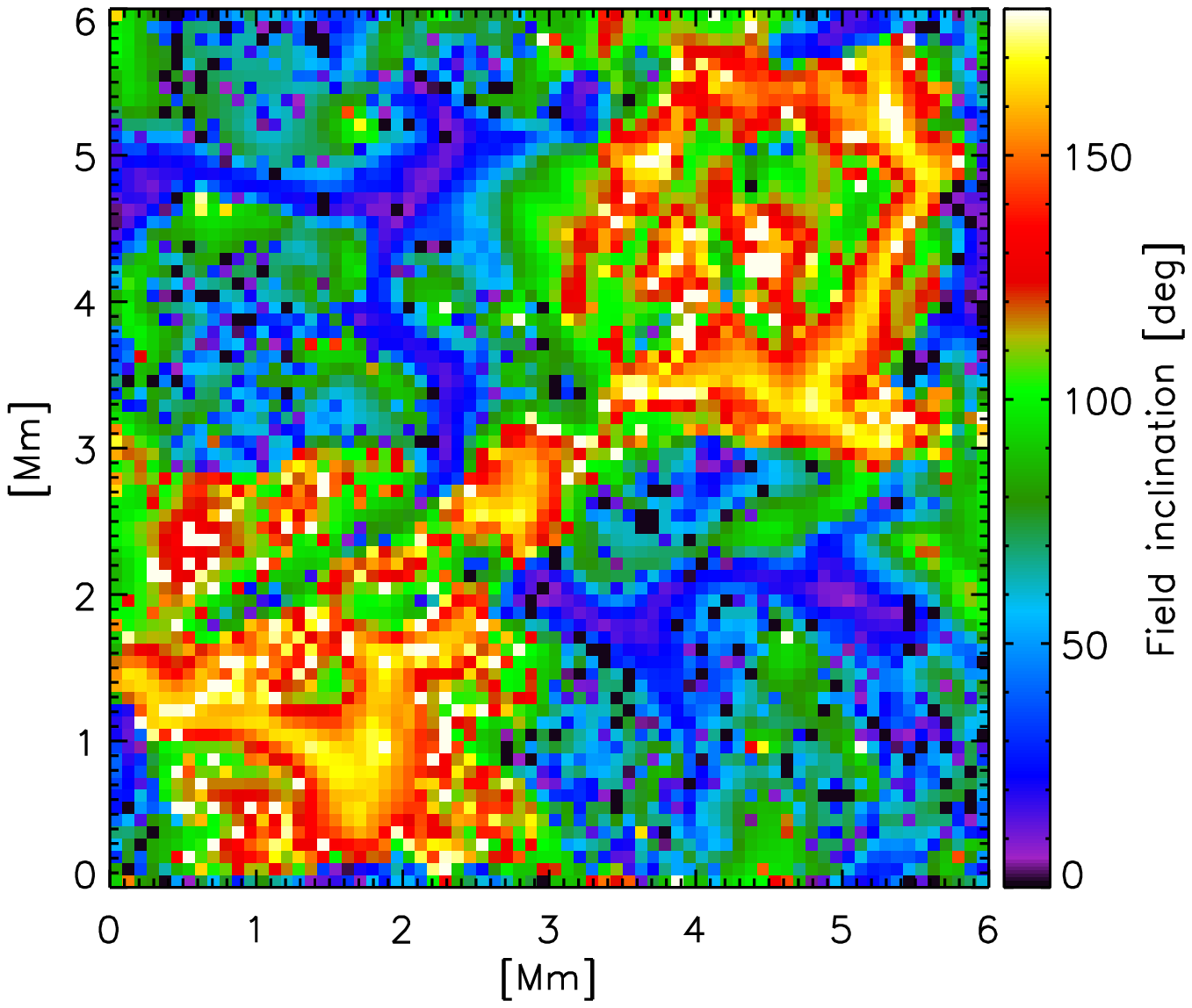}}
\resizebox{0.246\hsize}{!}{\includegraphics[trim=25mm 3mm 6mm
0mm,clip]{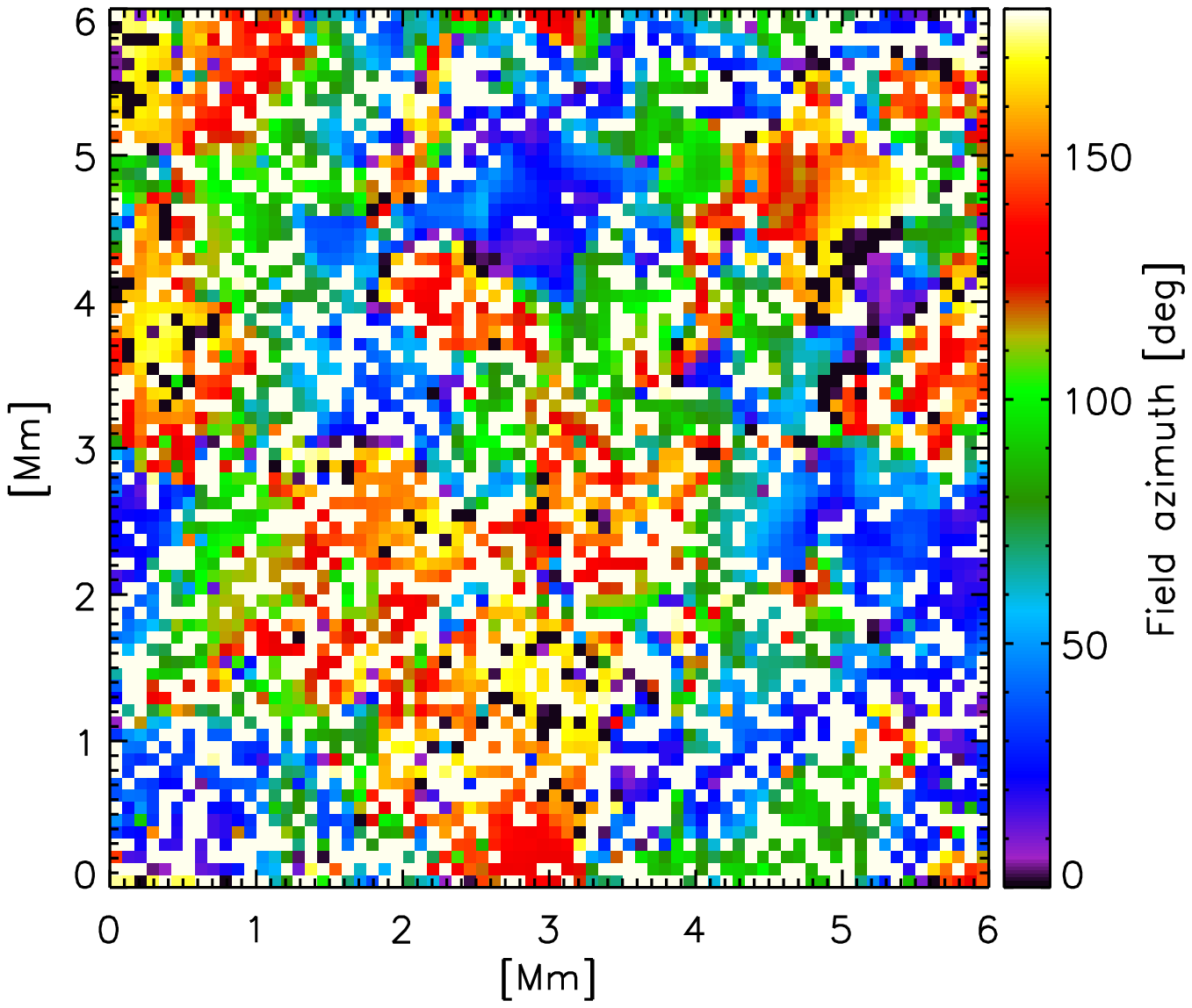}}
\resizebox{0.246\hsize}{!}{\includegraphics[trim=25mm 3mm 6mm
0mm,clip]{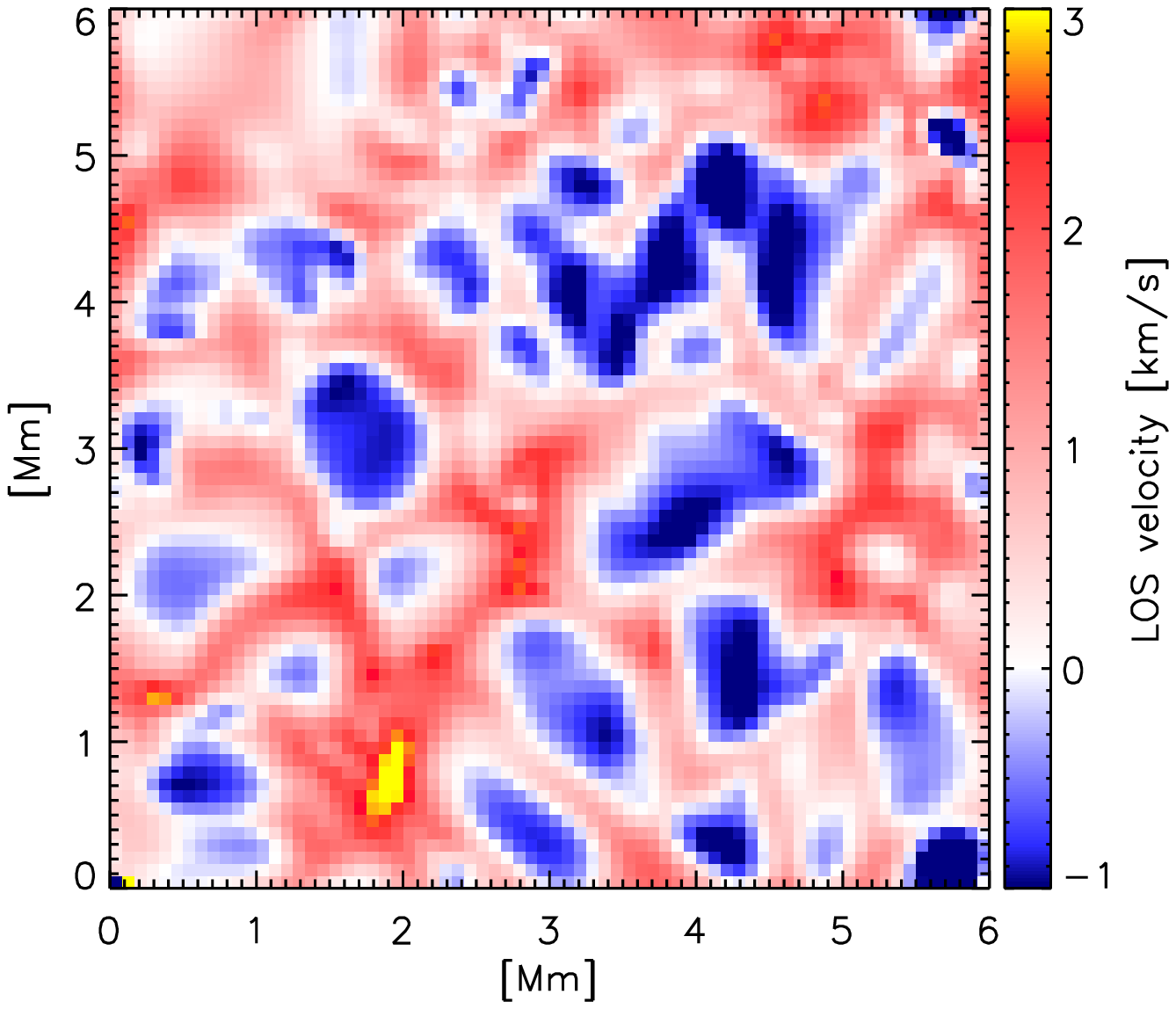}}
\resizebox{0.246\hsize}{!}{\includegraphics[trim=25mm 3mm 6mm
0mm,clip]{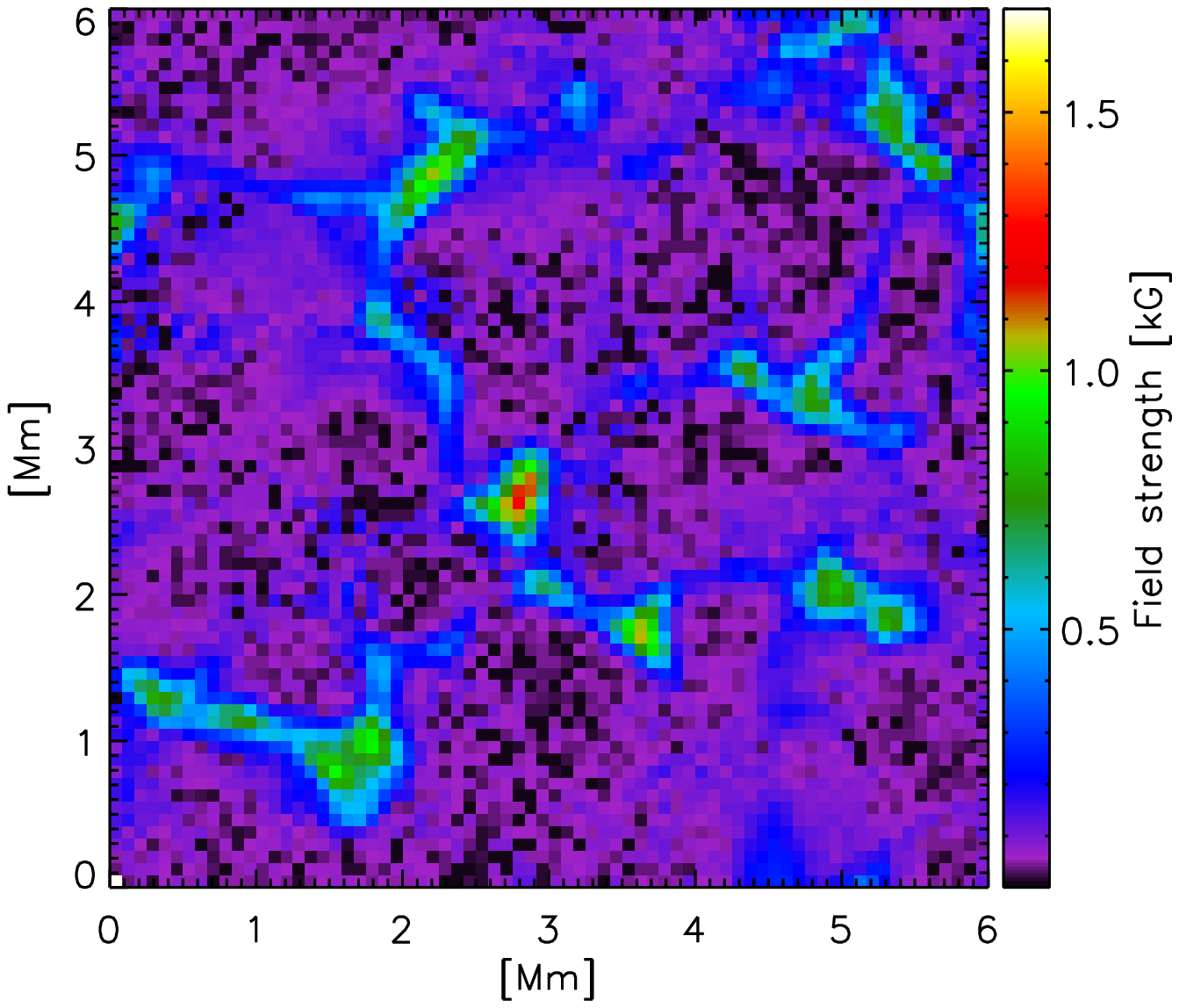}}
\resizebox{0.246\hsize}{!}{\includegraphics[trim=25mm 3mm 6mm
0mm,clip]{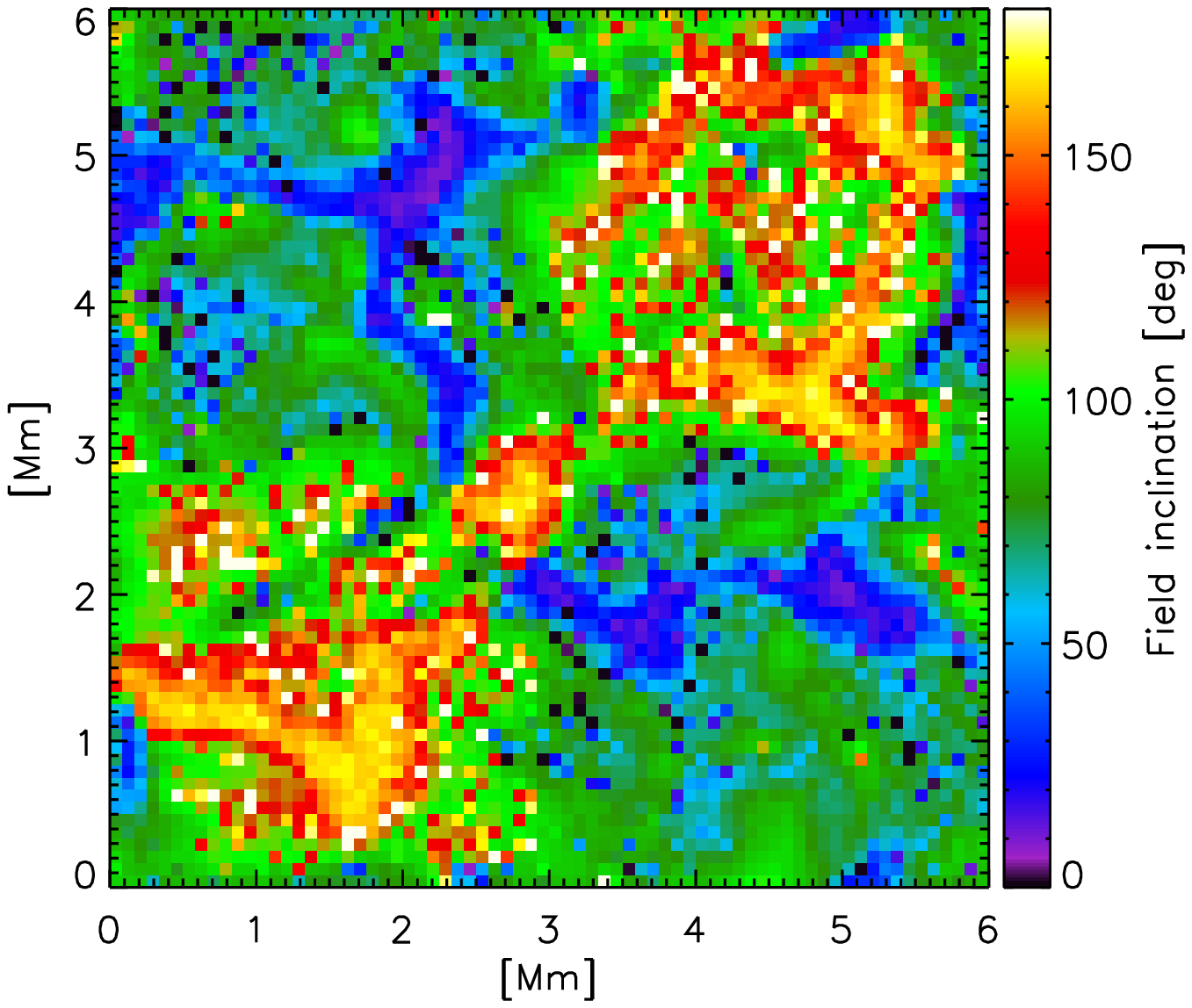}}
\resizebox{0.246\hsize}{!}{\includegraphics[trim=25mm 3mm 6mm
0mm,clip]{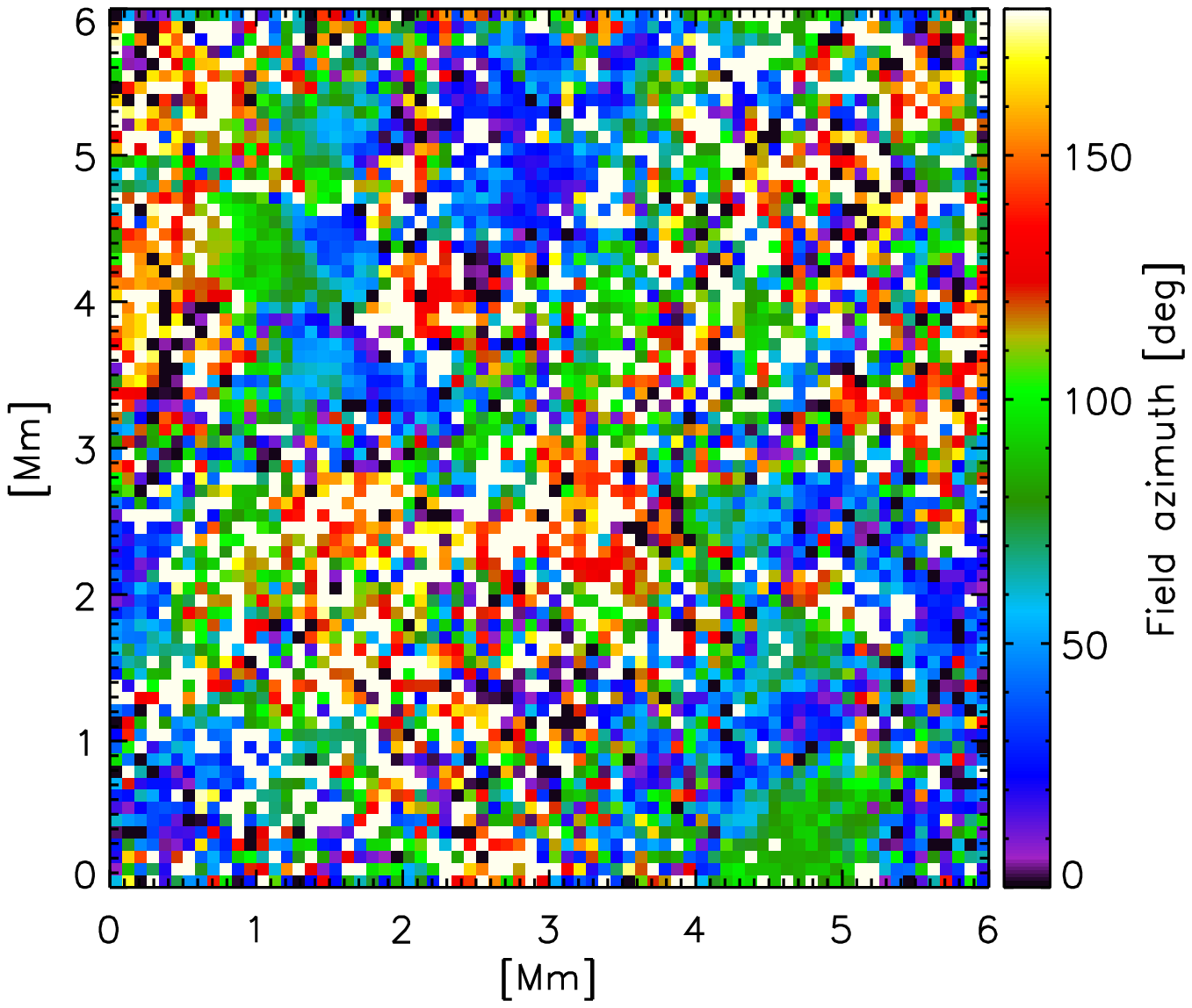}}
\resizebox{0.246\hsize}{!}{\includegraphics[trim=25mm 3mm 6mm
0mm,clip]{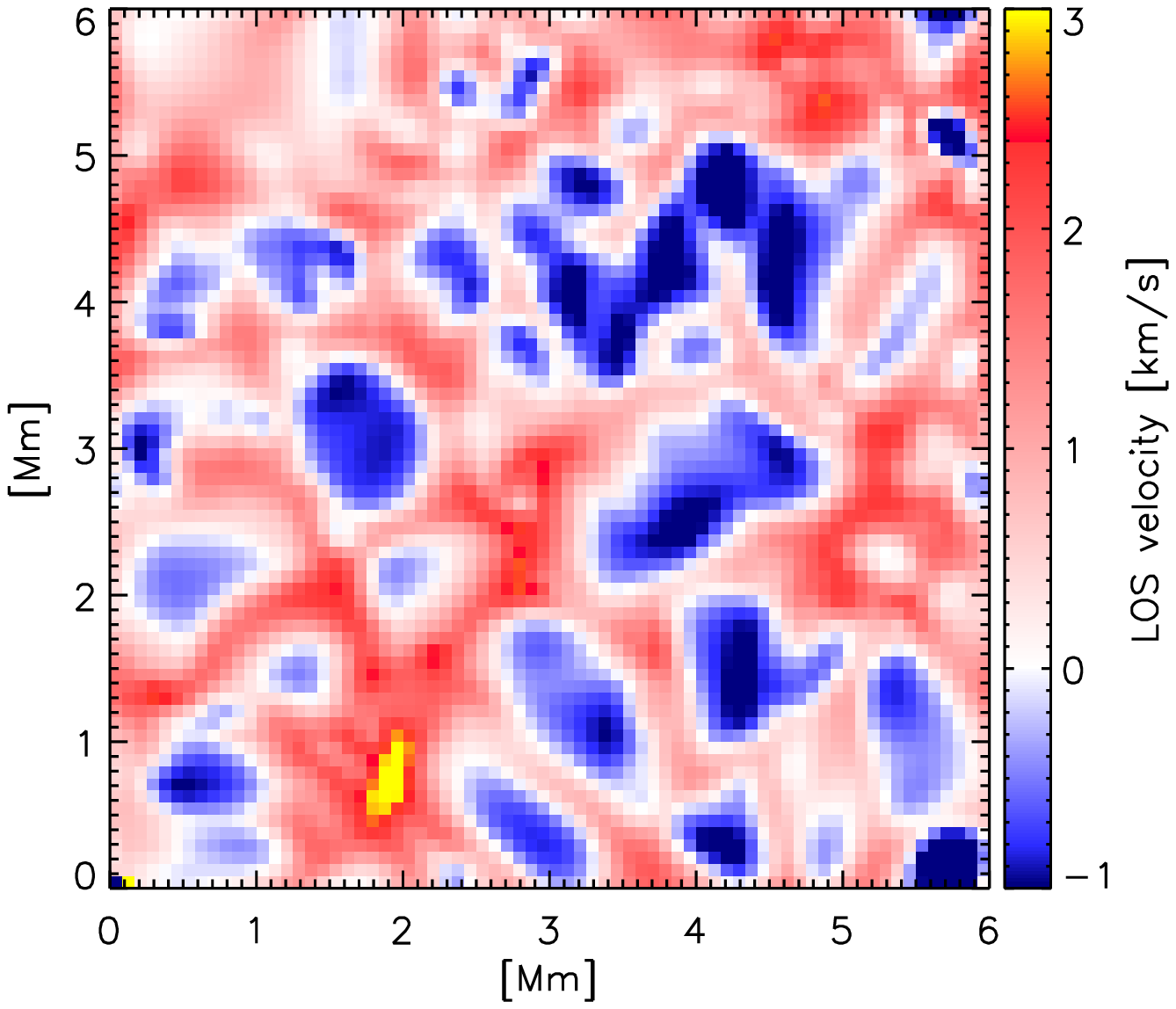}}
\resizebox{0.246\hsize}{!}{\includegraphics[trim=25mm 3mm 6mm 0mm,clip]{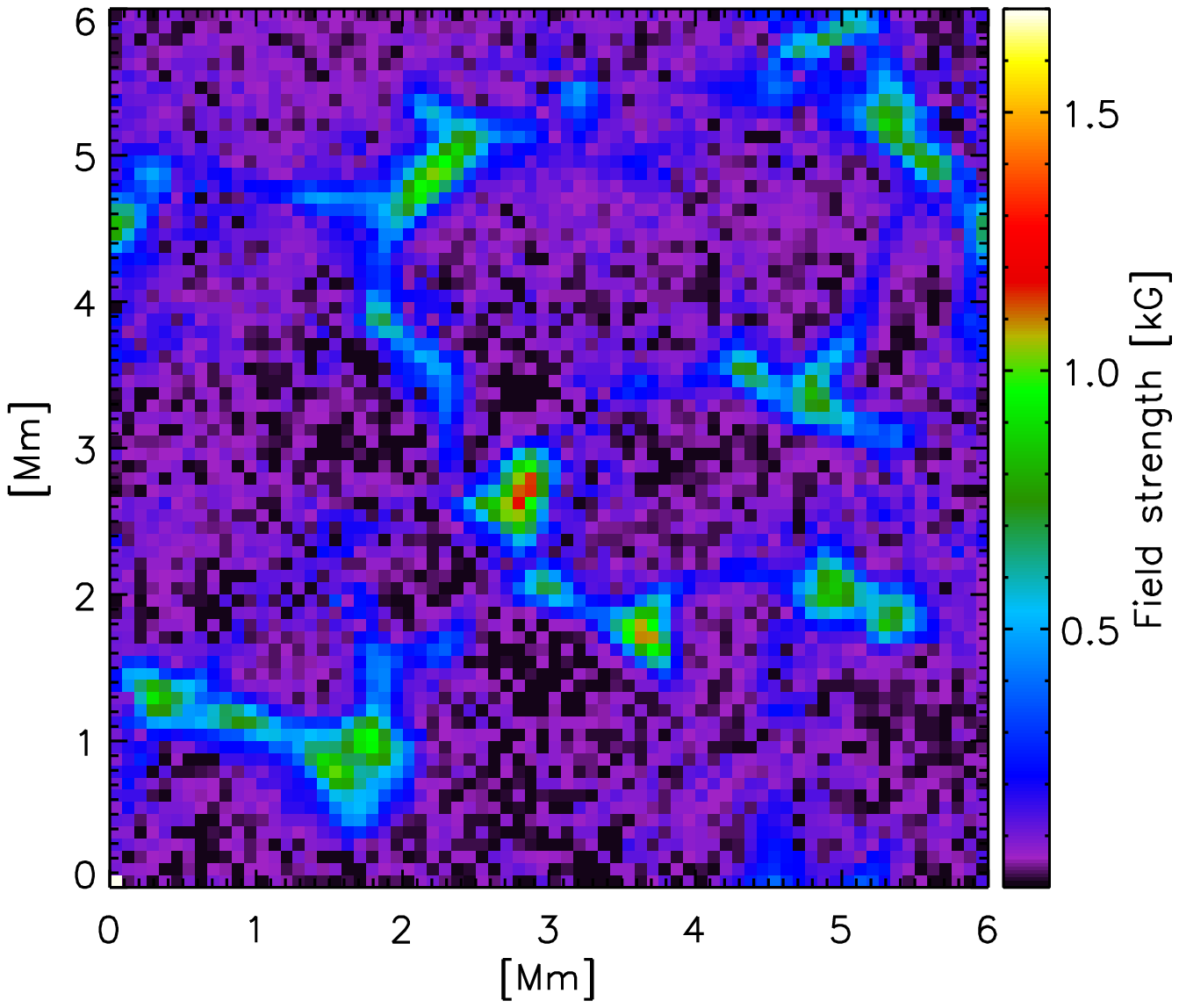}}
\resizebox{0.246\hsize}{!}{\includegraphics[trim=25mm 3mm 6mm 0mm,clip]{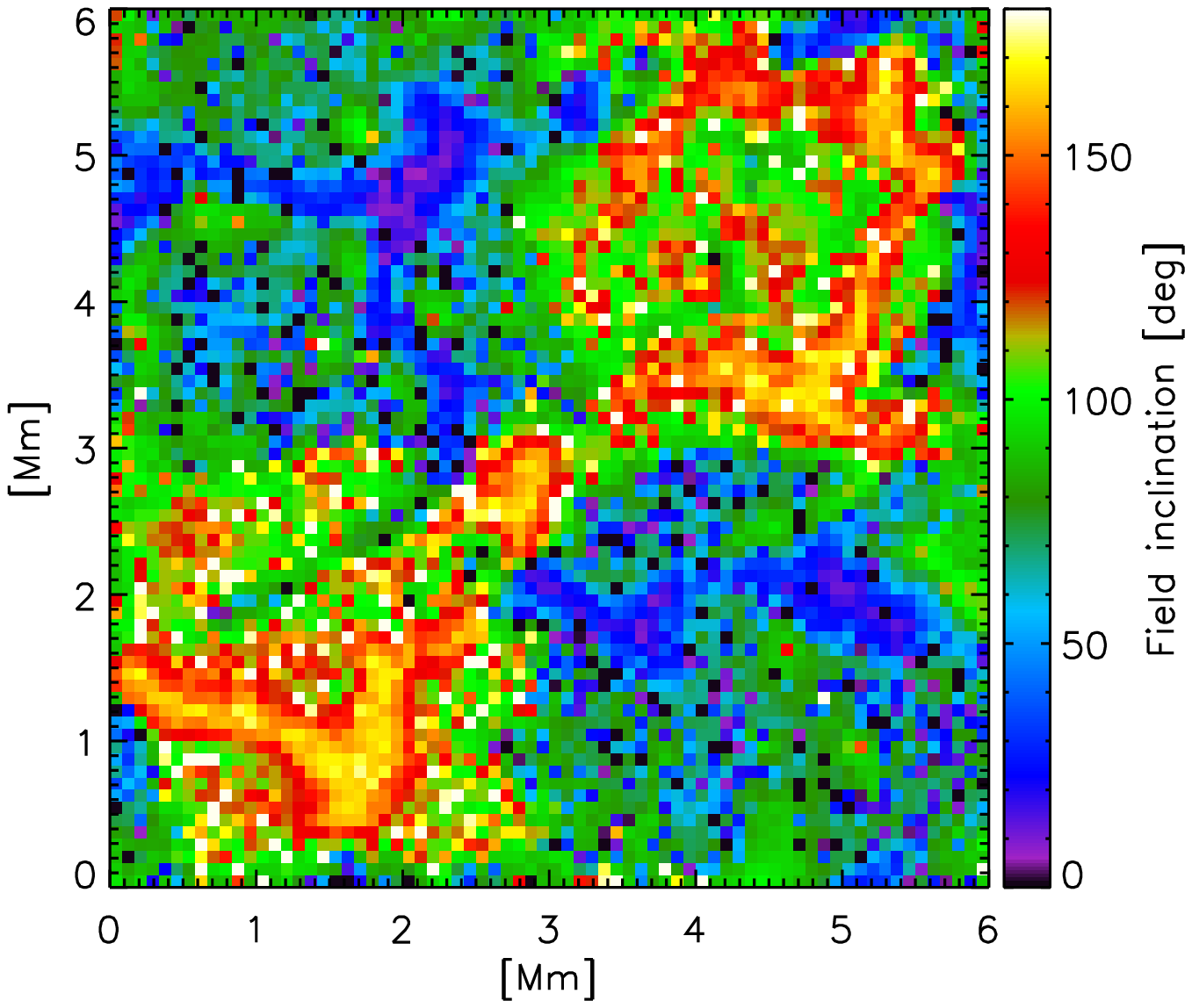}}
\resizebox{0.246\hsize}{!}{\includegraphics[trim=25mm 3mm 6mm 0mm,clip]{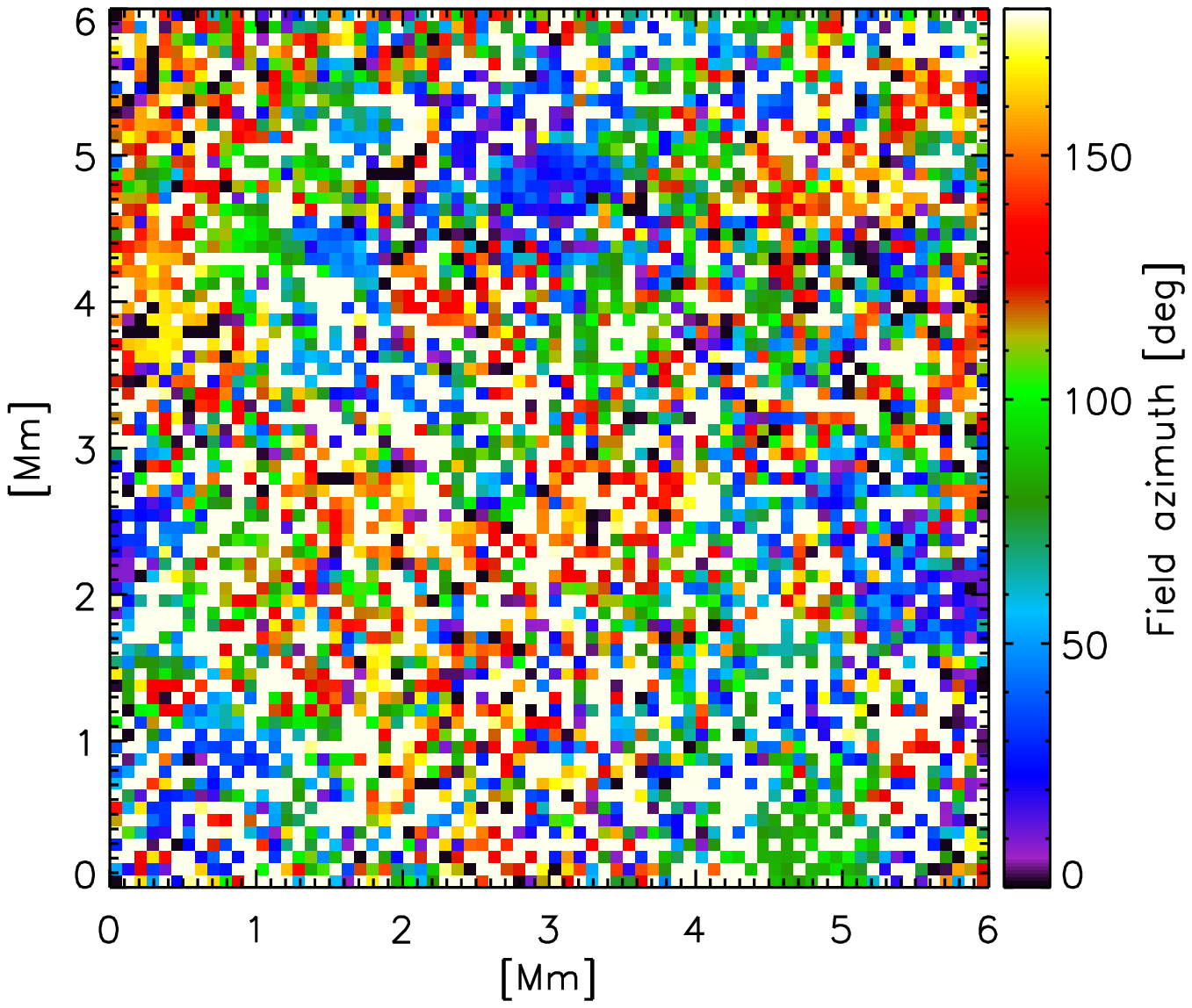}}
\resizebox{0.246\hsize}{!}{\includegraphics[trim=25mm 3mm 6mm 0mm,clip]{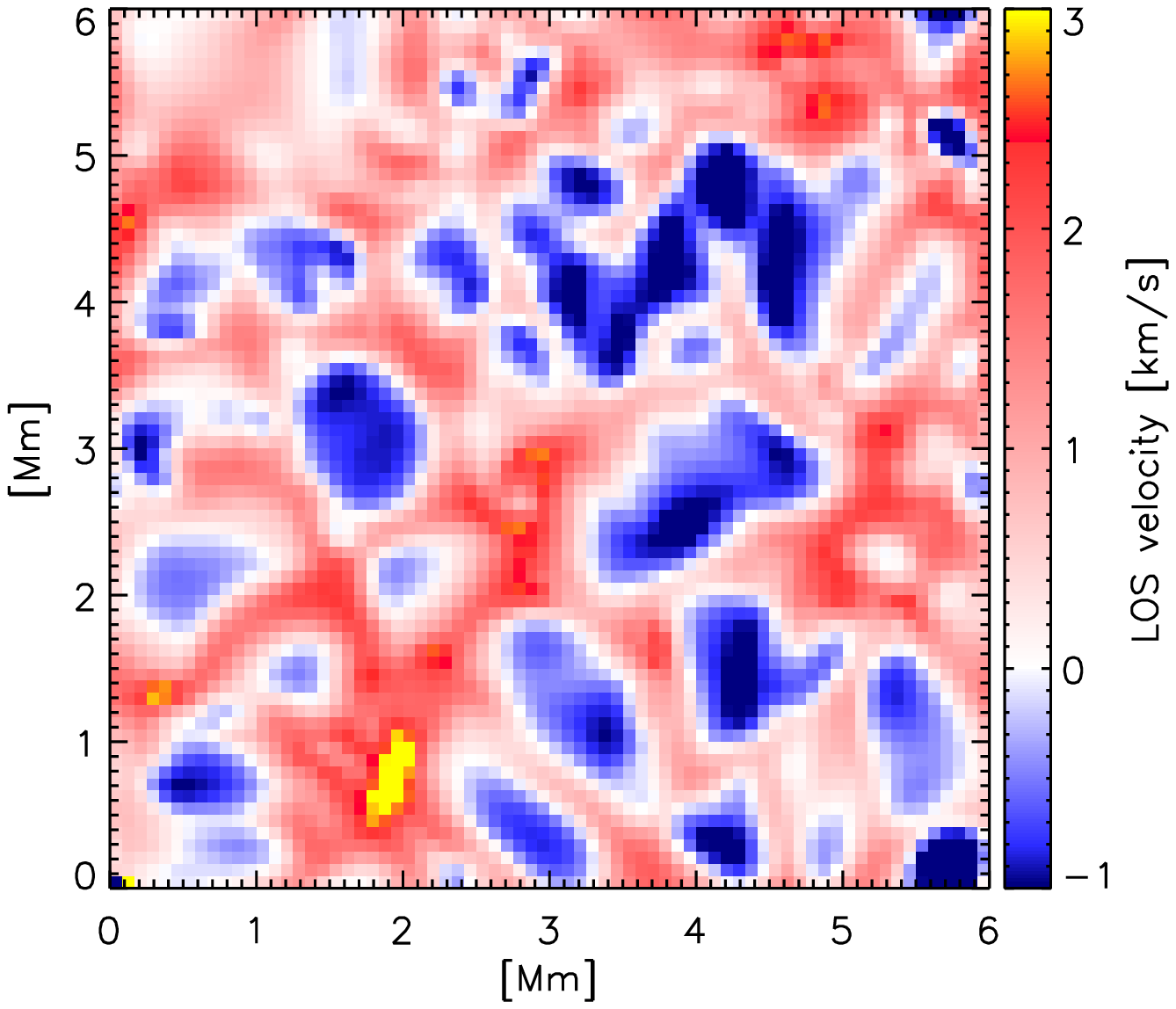}}
\caption{First row: Cut at optical depth $\mathrm{log}(\tau)=-1.5$ in 
the model atmospheres provided by the MHD simulations. Second row:
Maps of the physical quantities retrieved from the ME inversion of the
Stokes profiles with no noise, no spectral PSF, and 61 wavelength
samples.  Third row: Maps of the physical quantities retrieved from a
ME inversion of the Stokes profiles convolved with a 120 m{\AA} filter
and sampled at five wavelength points across the spectral line, plus a
continuum point. The SNR of the profiles is 1000. Fourth row: Same as
the third row, but for the profiles sampled at three wavelength
positions plus continuum. From left to right: magnetic field strength,
inclination, azimuth, and LOS velocity.\label{fig:mapas}}
\end{figure*}

Figure~\ref{fig:mapas} is a graphical illustration of the kind of
results we can expect from the analysis of VIM measurements. The upper
panels show a cut of the atmospheres provided by the MHD simulations
at optical depth $\mathrm{log}(\tau)=-1.5$. The second row display the
{\em reference} solution, i.e., the results of ME inversions of the
spatially degraded Stokes profiles with no noise, no spectral PSF, and
61 wavelength samples. The third row shows the atmospheric parameters
derived from the ME inversion in the specific case of five wavelength
samples at $-100$, $-50$, 0, 50, and 100~m{\AA} from line center plus
the continuum, a SNR of 1000, and an instrumental profile width of
120~m{\AA}. The last row shows the same parameters when the inversion
is applied to the Stokes profiles sampled at only three wavelength
positions ($-60$, $-10$, and 60~m{\AA}) plus the continuum, for a
filter width of 120~m{\AA} and a SNR of 1000.

The various physical parameters are qualitatively well determined,
although we observe some differences between the {\em real} and the
inferred parameters. The magnetic field strength, for example, is not
particularly well recovered inside the granules. There, the fields are
weak and the corresponding polarization signals are strongly affected
by the noise. In the inclination and azimuth maps we see regions fully
dominated by noise. In general, however, the inversion algorithm is
able to recover magnetic fields above 100~G with accuracy: {\em pixels
with weak fields are assigned weak fields, and pixels with strong
fields get strong fields}. This is in contrast with the results of
inversions of full Stokes profiles of internetwork fields in the quiet
Sun at resolutions of 1\arcsec (Mart\'{\i}nez Gonz\'alez et al.\
2006). Velocities are less affected by noise. We find larger velocity 
errors in intergranular regions, probably due to the larger asymmetries
exhibited by the Stokes profiles in those regions, where vertical
gradients are more pronounced.

\vspace{-1em}
\section{Conclusions}

We have analyzed simulated VIM-HRT observations to study the
performance of the instrument. Stokes profiles of the Fe~{\small
I}~617.3~nm line have been calculated using realistic MHD 
simulations of a quiet sun region at disk center and then spatially degraded 
by telescope diffraction and detector pixel size to match the 
VIM-HRT resolution. Additionally, we have convolved the profiles 
with spectral PSFs of different widths (from 0 to 200 m\AA\/),
added noise at the level of $10^{-3} I_{\rm c}$, and selected 
a few wavelength samples across the line. 

The instrumental filter width and the limited wavelength sampling
influence the determination of vector magnetic fields and LOS
velocities. However, the atmospheric parameters retrieved from VIM-HRT
measurements are reasonably accurate: ME inversions of the Stokes
profiles broadened by a 120~m\AA\/ filter and sampled at three
wavelength positions within the line plus a continuum point indicate
rms errors of $\sigma_{\rm B} \approx 30$~G, $\sigma_\mathrm{v}
\approx 50$~m/s and $\sigma_\mathrm{\gamma} \approx 10^\mathrm{o}$ 
for SNRs of 1000. As expected, the rms errors are much smaller for 
fields stronger than 750~G.

The results of ME inversions seem to be accurate enough even with
filters as wide as 120~m{\AA} FWHM and four wavelength samples. This
may allow both a significant reduction of the mass of VIM and better
signal-to-noise ratios through longer exposure times. There are some
drawbacks in using broad filters and few wavelength samples,
however. They include the decrease of the number of detectable
polarization signals and the reduction of the dynamical range the
instrument is able to cope with. Such effects are important and have
to be investigated in detail. We also note that the observation of
only four wavelength samples may prevent estimates of the magnetic
filling factor from being made, although our tests show that unity
filling factors produce good results with high spatial resolution
data.

\vspace{-1em}
\section*{acknowledgements}
\vspace{-1em}
This work has been partially funded by the Spanish Mi\-nisterio de
Educaci\'on y Ciencia through project ESP2003-07735-C04-03 (in\-clu\-ding
European FEDER funds) and {\em Programa Ram\'on y Cajal}.

\end{document}